\documentclass[11pt]{article}

\usepackage{graphicx,epsf, epsfig, amssymb, amsmath, amsfonts, mathrsfs}
\usepackage{longtable}
\usepackage{bm}
\usepackage{color}
\usepackage{enumitem}
\usepackage{float}
\usepackage{subcaption}
\usepackage{upgreek}
\usepackage{cite}
\usepackage{mathtools}
\usepackage{titlesec}
\usepackage{lipsum}
\usepackage{xcolor}
\usepackage{booktabs}
\usepackage{multirow}
\usepackage[section]{placeins}
\usepackage[colorlinks=true,linktocpage=true,linkcolor=blue,citecolor=blue]{hyperref}
\usepackage{tensor}
\usepackage[utf8]{inputenc}
\usepackage{authblk}
\usepackage{array}
\usepackage{breqn}
\usepackage{soul}

\newcommand{\be}{\begin{equation}}
\newcommand{\ee}{\end{equation}}
\newcommand{\bea}{\begin{eqnarray}}
\newcommand{\eea}{\end{eqnarray}}

\newcommand{\dmoff}[1]{}

\usepackage{caption}
\newsavebox\tmpbox

\title{\bf Relativistic and Newtonian Proca Stars:\\ A Tale of Two Limits}
\author[1]{Alberto Diez-Tejedor}
\author[2,3]{Carlos Herdeiro}
\author[4]{Claudio Lazarte}
\author[2]{Eugen Radu}
\author[5]{Armando A. Roque}
\author[4]{Nicolas Sanchis-Gual}
\author[2,6]{Etevaldo dos Santos Costa Filho}

\affil[1]{Departamento de Física, División de Ciencias e Ingenierías, Campus León, Universidad de Guanajuato, 37150, León, México}
\affil[2]{Departamento de Matemática da Universidade de Aveiro and Centre for Research and Development
in Mathematics and Applications (CIDMA), Campus de Santiago, 3810-193 Aveiro, Portugal}
\affil[3]{Programa de P\'os-Gradua\c{c}\~{a}o em F\'{\i}sica, Universidade Federal do Par\'a, 66075-110, Bel\'em, Par\'a, Brazil}
\affil[4]{Departament d'Astronomia i Astrof\'isica, Universitat de Val\`encia, Avinguda Vicent Andr\'es Estell\'es 19, 46100 Burjassot (Val\`encia), Spain}
\affil[5]{Unidad Acad\'emica de F\'isica, Universidad Aut\'onoma de Zacatecas, 98060 Zacatecas, M\'exico}
\affil[6]{Programa de Pós-Graduação em Física, Universidade Federal do Espírito Santo, Vitória, ES,  29075-910, Brazil}

\begin{document}

\maketitle

\begin{abstract}
We investigate a representative set of static solitonic solutions of the Einstein-Proca theory in the Newtonian regime, where the field frequency approaches the particle mass, $\omega \to \mu$, and compare them with the corresponding solutions of the spin-1 Schr\"odinger-Poisson system, which provides the effective description in this limit. While this correspondence is relatively straightforward in the Einstein-Klein-Gordon case, the vector nature of the Proca field and the emergence of an enhanced $U(3)$ symmetry in the nonrelativistic spin-1 limit give rise to several nontrivial features that require careful analysis. We establish a mapping between the two descriptions by identifying $\ell=0$ electric Proca stars with radially polarized (hedgehog) configurations and $\ell=1$ electric Proca stars with linearly polarized  configurations. We further clarify some aspects of the ground state and resolve several apparent discrepancies between  relativistic and Newtonian solutions, particularly concerning their morphology and stability properties. An important conclusion of this work is that the nonrelativistic regime supports a richer spectrum of stable equilibrium configurations than the relativistic theory, including stable excited states.
\end{abstract}

\newpage

\begingroup
\linespread{1.0}\selectfont
\tableofcontents
\endgroup

\newpage

%%%%%%%%%%%%%%%%%%%%%%%%%%%%%%%%%%%%%%%%%%%%%%%%%%%
\section{Introduction}\label{sec:1}
%%%%%%%%%%%%%%%%%%%%%%%%%%%%%%%%%%%%%%%%%%%%%%%%%%%

Proca stars are localized classical solutions of the Einstein-Proca equations that are regular everywhere and decay sufficiently rapidly at spatial infinity, resulting in configurations with finite total energy. They were first constructed in the fully relativistic framework in Ref.~\cite{Brito:2015pxa}, and have since been extensively studied in the literature (see, for example, Refs.~\cite{SalazarLandea:2016bys, Brihaye:2017inn, Sanchis-Gual:2017bhw, Minamitsuji:2018kof, Sanchis-Gual:2018oui, Herdeiro:2020kba, Herdeiro:2020jzx, CalderonBustillo:2020fyi, Herdeiro:2021lwl, Sanchis-Gual:2022mkk, Rosa:2022tfv, Herdeiro:2023wqf, Wang:2023tly, Lazarte:2024jyr, Sengo:2024pwk, Joaquin:2024quo, Herdeiro:2023lze, Herdeiro:2024pmv, Mio:2026wfh, Herdeiro:2026agu}). The relativistic description captures the full spectrum of solitonic configurations supported by the Einstein-Proca system and is essential for investigating strongly self-gravitating compact objects~\cite{Cardoso:2019rvt}, as well as their possible gravitational wave signatures~\cite{CalderonBustillo:2020fyi, Sanchis-Gual:2022mkk, Palloni:2025mhn}.

For many applications, however, such a fully relativistic description is not required. In particular, when considering phenomena such as galactic dynamics in models of ultralight dark matter~\cite{Marsh:2015xka,Hui:2021tkt,Ferreira:2020fam}, the relevant configurations typically lie well within the nonrelativistic regime. In this limit, the system can be accurately described within the Newtonian approximation, which consists of the \mbox{spin-1}  Schr\"odinger-Poisson equations. This framework is significantly simpler and computationally more efficient, while still capturing the essential physics. Proca stars in this regime have been investigated in Refs.~\cite{Jain:2021pnk, Zhang:2021xxa, Adshead:2021kvl, Gorghetto:2022sue, Zhang:2023ktk, Chen:2024vgh, Zhang:2024bjo, Nambo:2024hao, Nambo:2025lnu, Diaz-Andrade:2026hgn, Schiappacasse:2026ems}.

However, a quick comparison between the results reported for relativistic and Newtonian Proca stars appears to reveal some  inconsistencies (see Table~\ref{table}). For instance, the ground state of the Newtonian theory~\cite{Nambo:2024hao}, defined as the configuration of lowest energy at fixed particle number, is spherically symmetric, whereas in the relativistic theory it has been argued to be prolate~\cite{Herdeiro:2023wqf}. Furthermore, nodeless radially polarized states, which constitute stable excited solutions in the Newtonian theory~\cite{Nambo:2025lnu}, have recently been shown to be unstable  against generic non-spherical perturbations in the relativistic case~\cite{Herdeiro:2023wqf}. The main purpose of this paper is to clarify these puzzles and to show that, despite these apparent discrepancies, the relativistic and Newtonian analyses are in fact fully consistent.

We address this issue by explicitly exploring the Newtonian limit of the relativistic solutions and comparing them with the solutions obtained directly from the Newtonian limit of the Einstein-Proca theory.
In particular, we show that $\ell=0$ electric Proca stars~\cite{Brito:2015pxa} reduce, in the limit $\omega \to \mu$, to radially polarized  Proca stars~\cite{Nambo:2024hao}, whereas $\ell=1$ electric Proca stars~\cite{Herdeiro:2023wqf} reduce, in the same limit, to linearly polarized Proca stars~\cite{Nambo:2024hao}. This correspondence indicates that both the non-spherical structure of the ground state energy density and the instability of nodeless radially polarized configurations are intrinsically relativistic features that disappear in the Newtonian regime, even though for the ground state the Proca field still retains a preferred polarization direction. This interpretation is further supported by fully nonlinear numerical evolutions, which show that the lifetime of radially polarized configurations rapidly increases as $\omega\to\mu$, consistent with the onset of stability in the Newtonian limit.

This paper is organized as follows. In Sec.~\ref{sec:2}, we introduce the minimal theoretical framework required for our analysis, including the Einstein-Proca theory and its Newtonian limit, described by the spin-$1$ Schr\"odinger-Poisson  system. In this section we also present the ans\"atze for the most relevant configurations in each regime, namely the monopolar ($\ell=0$) and dipolar ($\ell =1$) electric solutions in the relativistic theory, and the radial (hedgehog) and linear polarization states in the Newtonian limit. Although this selection of configurations may appear somewhat arbitrary, it encompasses all static solutions currently known to be dynamically stable, excluding multi-frequency states~\cite{Nambo:2024hao,Diaz-Andrade:2026hgn}, whose existence appears to be tied to the enhanced $U(3)$ symmetry that emerges in the nonrelativistic limit and which therefore do not have a clear relativistic analogue. In Sec.~\ref{sec.numerical}, we construct both relativistic and Newtonian Proca star solutions numerically. Section~\ref{sec.relativistic_newtonian_conection} contains the main results of this work, where we perform a detailed comparison between these configurations and establish a dictionary that relates solutions across the two regimes. In particular, we demonstrate full agreement between the relativistic and Newtonian descriptions in the regime where both are applicable, showing that, as $\omega\to\mu$, the $\ell=1$ electric Proca stars progressively lose their prolate deformation, while the instability timescale of the $\ell=0$ electric Proca stars becomes arbitrarily large. Although the existence of a correspondence is physically expected, it is not entirely trivial, since solutions of the full theory need not necessarily reduce, in the appropriate limit, to solutions of the limiting theory, nor vice versa. In fact, we find that the Newtonian regime admits a richer spectrum of stable equilibrium configurations, which we interpret as a consequence of the symmetry enhancement of the nonrelativistic theory. We conclude in Sec.~\ref{sec.conclusions} with a brief summary of our findings.

\begin{table}[t!]
\centering
\begin{tabular}{lcc|lcc}
\toprule
\multicolumn{3}{c|}{Relativistic} & \multicolumn{3}{c}{Newtonian} \\
\cmidrule(lr){1-3}\cmidrule(lr){4-6}
\multicolumn{1}{c}{State} & \multicolumn{1}{c}{Morphology} & \multicolumn{1}{c|}{Stable} &
\multicolumn{1}{c}{State} & \multicolumn{1}{c}{Morphology} & \multicolumn{1}{c}{Stable} \\
\midrule
$\ell=0\ \text{electric}$  & Spherical  & \textcolor{red!70!black}{No} 
& Radial polarization   & Spherical  & \textcolor{red!70!black}{$n=0$} \\
$\ell=1\ \text{electric}$  & \textcolor{green!50!black}{Prolate} & $n=0^*$  
& Linear polarization   & \textcolor{green!50!black}{Spherical} & $n=0$ \\
\multicolumn{3}{c|}{No correspondence} 
& Multi-frequency & Spherical & Bands \\
\bottomrule
\end{tabular}
\caption{\small {\bf Correspondence between relativistic and Newtonian Proca stars.} States appearing in the same row represent the same physical configuration in the two regimes of the theory. A naive comparison of properties reported in the relativistic and Newtonian literature suggests some apparent discrepancies, highlighted by matching colors, which are clarified in this work. The star denotes families for which only a subset of the states are stable.}
\label{table}
\end{table}

%%%%%%%%%%%%%%%%%%%%%%%%%%%%%%%%%%%%%%%%%%%%%%%%%%%
\section{Theoretical framework}\label{sec:2}
%%%%%%%%%%%%%%%%%%%%%%%%%%%%%%%%%%%%%%%%%%%%%%%%%%%

In this section, we introduce the minimal theoretical background required to follow this paper. In particular, we present the Einstein-Proca theory, its Newtonian limit (which is described by the 
\mbox{spin-1} Schr\"odinger-Poisson equations), and the ans\"atze for the solutions that we explore here. Further details can be found in the original references.

%%%%%%%%%%%%%%%%%%%%%%%%%%%%%%%%%%%%%%%%%%%%%%%%%%%
\subsection{Einstein-Proca theory}\label{sec:2.1}
%%%%%%%%%%%%%%%%%%%%%%%%%%%%%%%%%%%%%%%%%%%%%%%%%%%

Our starting point is the action of a free complex Proca field $\mathcal{A}_\alpha$ minimally coupled to Einstein's gravity:\footnote{Note that the Proca action used in~\cite{Brito:2015pxa} differs by an overall factor of 2 from that in~\cite{Nambo:2024hao}. At the classical level, this difference amounts only to a choice of normalization, and the two conventions are therefore equivalent. In this paper we adopt the normalization of the former.\label{footnote}}
\begin{equation}
\label{eq.EP.action}
\mathcal{S}[g_{\alpha\beta},\mathcal{A}_\alpha,\mathcal{A}^{*}_{\alpha}] = \int\text{d}^4x\sqrt{-g}\left(\frac{R}{16\pi G}-\frac{1}{4}F_{\alpha\beta}^*F^{\alpha\beta}-\frac{1}{2}\mu^2\mathcal{A}^*_\alpha\mathcal{A}^\alpha\right) ,
\end{equation}
where $G$ is Newton's gravitational constant, $g_{\alpha\beta}$ is the spacetime metric, with determinant $g$ and Ricci scalar $R$, $F_{\alpha\beta}:=\nabla_\alpha\mathcal{A}_\beta-\nabla_\beta\mathcal{A}_\alpha$ is the field strength tensor associated with the Proca field, $\mu$ denotes its bare mass, and a star will hereafter indicate complex conjugation.

Varying the action~\eqref{eq.EP.action} with respect to the vector field and the metric tensor yields, respectively, the Proca and  Einstein equations,
\begin{subequations}\label{eq.EP} 
\begin{eqnarray}
&\nabla_\alpha F^{\alpha\beta}-\mu^2\mathcal{A}^\beta=0,\label{eq:2.2b}\\
&\displaystyle{R_{\alpha\beta}-\frac{1}{2}g_{\alpha\beta}R=8\pi GT_{\alpha\beta}}, \label{eq:2.2a}
\end{eqnarray}
\end{subequations}
where
\begin{align}\label{eq:2.3}
T_{\alpha\beta}=\frac{1}{2}g^{\sigma\gamma}\left(F_{\alpha\sigma}F^*_{\beta\gamma}+F^*_{\alpha\sigma}F_{\beta\gamma}\right)-\frac{1}{4}g_{\alpha\beta}F^*_{\sigma\gamma}F^{\sigma\gamma}+\frac{1}{2}\mu^2\left(\mathcal{A}_\alpha\mathcal{A}^*_\beta+\mathcal{A}^*_\alpha \mathcal{A}_\beta-g_{\alpha\beta}\mathcal{A}^*_\sigma\mathcal{A}^\sigma\right)
\end{align}
is the stress-energy tensor of the vector field. Taking the covariant divergence of Eq.~\eqref{eq:2.2b}, and using the antisymmetry of $F^{\alpha\beta}$, one obtains the Lorenz condition as a dynamical constraint,
\begin{equation}\label{eq:lorenz}
\nabla_\alpha \mathcal{A}^\alpha=0.
\end{equation}
In contrast to Maxwell theory, this relation should not be interpreted as a gauge choice, since the mass term explicitly breaks gauge invariance. Instead, it follows directly from the equations of motion and enforces the correct number of propagating degrees of freedom for a massive spin-1 field. In particular, it implies that the temporal component of the vector field, $\mathcal{A}_0$, is not an independent dynamical variable, while the three physical propagating degrees of freedom are encoded in the spatial components $\mathcal{A}_i$, corresponding to the three polarization states of the field.

In addition, the action~\eqref{eq.EP.action} is invariant under global $U(1)$ transformations of the form $\mathcal{A}_\alpha\mapsto e^{i\chi}\mathcal{A}_\alpha$, where $\chi$ is a real constant. By Noether's theorem, this symmetry implies the existence of the conserved four-current
\begin{align}
j^\alpha=\frac{i}{2}\left(F^{\alpha\beta *}\mathcal{A}_\beta-F^{\alpha\beta}\mathcal{A}^*_\beta\right),
\end{align}
which satisfies $\nabla_\alpha j^\alpha=0$. The associated conserved Noether charge is obtained by integrating the temporal component of the four-current over a spacelike hypersurface $\Sigma$, 
\begin{align}\label{eq.Q}
N=\int_\Sigma\text{d}^3x\sqrt{-g}j^0 .
\end{align}
In the quantum theory, this quantity corresponds to the total particle number carried by the Proca field, or more precisely to the net particle number, i.e. the difference between particles and antiparticles, and we will adopt this interpretation throughout this work.

For asymptotically flat spacetimes, the appropriate notion of total energy is provided by the ADM mass $M$, which can be interpreted as the conserved charge associated with asymptotic time translations at spatial infinity. In practice, this quantity can be extracted from the asymptotic behavior of the metric. In asymptotically Cartesian coordinates, the time-time component of the metric satisfies
\begin{equation}
g_{00} = -1 + \frac{2GM}{r} + \mathcal{O}(r^{-2}),
\end{equation}
independently of whether the spacetime is spherically symmetric, from which it follows that
\begin{equation}
M = \frac{1}{2G}\lim_{r\to\infty}r(1+g_{00}).
\end{equation}
Throughout this work, we compute the ADM mass from the asymptotic falloff of the metric functions.

For configurations containing only particles or only antiparticles, the ADM mass and the Noether charge allow us to define the binding energy as
\begin{equation}\label{eq.binding.energy}
E_{\rm bind}\equiv M-\mu |N|,
\end{equation}
where $\mu |N|$ represents the rest-mass energy of the constituent particles. Thus, $E_{\rm bind}$ measures the difference between the total energy of the self-gravitating configuration and the rest-mass energy of the corresponding collection of free particles at rest and at infinite separation. For a bound configuration, $E_{\rm bind}<0$, with $|E_{\rm bind}|$ representing the energy required to disperse the system into free particles at rest at infinity.

%%%%%%%%%%%%%%%%%%%%%%%%%%%%%%%%%%%%%%%%%%%%%%%%%%%
\subsection{Relativistic solitons and boundary conditions}\label{sec:2.2}
%%%%%%%%%%%%%%%%%%%%%%%%%%%%%%%%%%%%%%%%%%%%%%%%%%%

We are interested in solitonic solutions of the Einstein-Proca theory. For the purposes of the present work, it suffices to consider static and axisymmetric configurations. Spacetimes with these symmetries admit two commuting Killing vector fields, $\{\xi,\eta\}$, associated with time translations and rotations around a symmetry axis, respectively. Accordingly, one can introduce coordinates $(t,\varphi)$ adapted to these symmetries, such that
\begin{equation}
\xi=\partial_t, \quad \eta=\partial_\varphi.
\end{equation}
We further assume that the spacetime admits a two-dimensional submanifold orthogonal to $\{\xi,\eta\}$, on which spherical-like coordinates $(r,\theta)$ can be introduced so that the radial coordinate $r$ is always orthogonal to the polar coordinate $\theta$, i.e. $g_{r\theta}=0$. A line element compatible with the previous assumptions reads
\begin{align}\label{eq.metric}
\text{d}s^2=-e^{2F_0(r,\theta)}\text{d}t^2+e^{2F_1(r,\theta)}\left(\text{d}r^2+r^2\text{d}\theta^2\right)+e^{2F_2(r,\theta)}r^2\sin^2\theta\text{d}\varphi^2 ,
\end{align}
where $t\in(-\infty,+\infty)$, $r\in[0,+\infty)$, $\theta\in[0,\pi]$, $\varphi\in[0,2\pi)$, and $\{F_0,F_1,F_2\}$ are real functions of the spherical-like coordinates $(r,\theta)$.

In the following, we consider two classes of static and axisymmetric solitonic solutions belonging to the family commonly referred to as electric Proca stars. We focus on this family because, to the best of our knowledge, it contains the only static solutions of the Einstein-Proca equations currently known to be dynamically stable. These configurations are characterized by a nonvanishing temporal component of the Proca potential, hence  ``electric.''\footnote{Magnetic Proca stars, for which the temporal component of the Proca field vanishes, have recently been investigated in Ref.~\cite{Herdeiro:2026agu}, where they were shown to be dynamically unstable.} 
However, once the standard harmonic time dependence of the field is assumed, the Proca equations imply that an axially symmetric electric configuration necessarily also involves two magnetic potentials, $\mathcal{A}_r$ and $\mathcal{A}_\theta$. Accordingly, the most general electric ansatz compatible with axial symmetry takes the form 
\begin{align}\label{eq.ansatz.electric}
\mathcal{A}=e^{-i\omega t}\left(iV(r,\theta)\text{d}t+\frac{H_1(r,\theta)}{r}\text{d}r+H_2(r,\theta)\text{d}\theta\right),
\end{align}
where $\omega$ is the real oscillation frequency associated with the internal phase rotation of the field, which we take to be positive,\footnote{Solutions with negative frequency are related to positive-frequency solutions by complex conjugation of the Proca field and carry opposite Noether charge. In the quantum interpretation, the two branches correspond to configurations dominated by particles and antiparticles, respectively. Restricting to $\omega>0$ therefore entails no loss of generality.} and $\{V,H_1,H_2\}$ are real functions of the spherical-like coordinates $(r,\theta)$. Note that, although the field exhibits harmonic time dependence through the phase factor $e^{-i\omega t}$, observables, such as the stress-energy tensor and the Noether current $j^\alpha$, remain time independent and, consequently, the spacetime geometry may consistently be taken to be static. 

\begin{figure}[t!]
\centering
\includegraphics[width=0.7\textwidth]{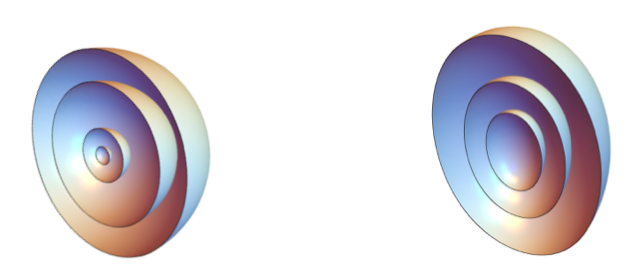}\\
$\ell=0$ electric \hspace{5.2cm} $\ell=1$ electric
\caption{\small\textbf{Surfaces of constant energy density.} {\it Left panel:}  Monopolar ($\ell = 0$) electric Proca star with $\omega = 0.980$. {\it Right panel:} Dipolar ($\ell= 1$) electric Proca star with $\omega =0.800$. Unlike the $\ell=0$ configuration, the $\ell=1$ solution is not spherically symmetric and exhibits a characteristic prolate deformation. }\label{fig.isosurfaces}
\end{figure}

Electric Proca stars admit different classes characterized by their angular structure, labeled by the integer $\ell = 0,1,2,\ldots$, corresponding to monopolar, dipolar, and higher multipolar configurations. This integer determines the angular dependence of the Proca field on the polar coordinate $\theta$ and, consequently, the geometric structure of the corresponding solitonic solution. In particular, under the reflection $\theta \to \pi-\theta$ across the equatorial plane, the Proca field exhibits a definite parity determined by $\ell$, with even and odd values of $\ell$ corresponding to even and odd parity, respectively. The spacetime metric, however, remains invariant under this reflection. 
In this work, we restrict our attention to the two lowest multipoles, namely $\ell=0$ and $\ell=1$, which are the configurations relevant for the comparison with the Newtonian solutions discussed later. Figure~\ref{fig.isosurfaces} shows representative surfaces of constant energy density for these two cases. In particular, note that the $\ell=1$ configurations exhibit a characteristic prolate deformation, in contrast to the spherically symmetric structure of the $\ell=0$ solutions.

%%%%%%%%%%%%%%%%%%%%%%%%%%%%%%%%%%%%%%%%%%%%%%%%
%\subsubsection{$\ell = 0$ electric Proca stars}
\subsubsection{\texorpdfstring{$\ell=0$}{l=0} electric Proca stars}
%%%%%%%%%%%%%%%%%%%%%%%%%%%%%%%%%%%%%%%%%%%%%%%%

The $\ell=0$ configurations correspond to monopolar solutions and were first constructed in Ref.~\cite{Brito:2015pxa}. In this case, the angular dependence becomes trivial and the system reduces to spherical symmetry. As a result, all fields depend only on the radial coordinate $r$, the two spatial metric functions coincide, $F_1(r)=F_2(r)$, the polar component of the vector field vanishes identically, $H_2(r)=0$, and the problem simplifies to a system of ordinary differential equations. (See Sec.~\ref{sec.numerical.rel} and App.~\ref{app.eqs} for more details.) 

The corresponding solutions are then determined by imposing regularity and asymptotic flatness.
At the origin, $r=0$, regularity requires the metric functions to be smooth and the Proca field to remain finite and well defined. In particular, this implies
\begin{subequations}\label{eq.boundary.ell=0.electric}
\begin{equation}
\left.\partial_r F_i(r)\right|_{r=0}=0,\quad \left.\partial_rV (r)\right|_{r=0}=0,\quad \left.H_1(r)\right|_{r=0}=0,
\end{equation}
which ensure the absence of singular behavior at the center. At spatial infinity, $r\to\infty$, on the other hand, asymptotic flatness requires the spacetime metric to approach Minkowski space and that the matter fields decay sufficiently rapidly so that the ADM mass and Noether charge remain finite. This leads to
\begin{equation}
\lim_{r\to\infty} F_i(r)=0,\quad \lim_{r\to\infty} V(r)=0,\quad \lim_{r\to\infty} H_1(r)=0 .
\end{equation}
\end{subequations}

These boundary conditions define regular, localized self-gravitating solitonic configurations, which we construct numerically in Sec.~\ref{sec.numerical.rel.ell=0}.

%%%%%%%%%%%%%%%%%%%%%%%%%%%%%%%%%%%%%%%%%%%%%%
%\subsubsection{$\ell= 1$ electric Proca stars}
\subsubsection{\texorpdfstring{$\ell=1$}{l=1} electric Proca stars}
%%%%%%%%%%%%%%%%%%%%%%%%%%%%%%%%%%%%%%%%%%%%%

The $\ell=1$ configurations correspond to dipolar solutions and were first constructed in Ref.~\cite{Herdeiro:2023wqf}. In contrast to the monopolar case, these solutions are intrinsically axisymmetric and exhibit a nontrivial angular dependence, reflecting their dipolar structure. As a result, the fields depend on both the radial and polar coordinates, $(r,\theta)$, and the problem must be solved as a system of coupled partial differential equations. (See Sec.~\ref{sec.numerical.rel} and App.~\ref{app.eqs} for  details.)

The solutions are determined by imposing regularity, asymptotic flatness, and axisymmetry. At the origin, $r=0$, regularity requires that all fields be smooth and free of singularities. This leads to the conditions
\begin{subequations}\label{eq.boundary.ell=1.electric}
\begin{equation}\label{eq.boundary.ell=1.electric1}
\left.\partial_r F_i(r,\theta)\right|_{r=0}=0,\quad \left.V(r,\theta)\right|_{r=0}=0,\quad \left.H_i(r,\theta)\right|_{r=0}=0,
\end{equation}
which ensure the absence of singular behavior at the center. At spatial infinity, $r \to \infty$, asymptotic flatness together with localization of the matter fields require
\begin{equation}\label{eq.boundary.ell=1.electric2}
\lim_{r\to\infty} F_i(r,\theta)=0,\quad \lim_{r\to\infty} V(r,\theta)=0,\quad \lim_{r\to\infty} H_i(r,\theta)=0,
\end{equation}
so that the spacetime approaches Minkowski space and the Proca field decays sufficiently fast for the total mass and Noether charge to remain finite. Along the symmetry axis, $\theta=0,\pi$, regularity and axisymmetry impose
\begin{equation}\label{eq.boundary.ell=1.electric3}
\left.\partial_\theta F_i(r,\theta)\right|_{\theta=0,\pi}=0,\quad \left.\partial_\theta V(r,\theta)\right|_{\theta=0,\pi}=0,\quad \left.\partial_\theta H_1(r,\theta)\right|_{\theta=0,\pi}=0,\quad \left.H_2(r,\theta)\right|_{\theta=0,\pi}=0,
\end{equation}
ensuring that the fields are well-defined and single-valued on the axis. Also, for all solutions, the metric functions and the Proca potential $H_2$ are symmetric with respect to a reflection on the equatorial plane, while $H_1$ and $V$ change sign as $\theta \to \pi -\theta$. As a result, it is enough to consider the range $0 \leq \theta  \leq \pi/2$ for the angular variable, with the following boundary conditions on the equatorial plane:
\begin{equation}\label{eq.boundary.ell=1.electric3n}
\left.\partial_\theta F_i(r,\theta)\right|_{\theta=\pi/2}=0,\quad 
\left. V(r,\theta)\right|_{\theta=\pi/2}=0,\quad \left. H_1(r,\theta)\right|_{\theta=\pi/2}=0,\quad \left.\partial_\theta H_2(r,\theta)\right|_{\theta=\pi/2}=0 .
\end{equation}
In our numerical implementation, the boundary conditions imposed explicitly are those given by Eqs.~(\ref{eq.boundary.ell=1.electric1}),~(\ref{eq.boundary.ell=1.electric2}), and~(\ref{eq.boundary.ell=1.electric3}).
In addition, the absence of conical singularities requires
\begin{equation}
\left.F_1(r,\theta)\right|_{\theta=0,\pi} = \left.F_2(r,\theta)\right|_{\theta=0,\pi},
\end{equation}
\end{subequations}
which is not imposed independently, but checked a posteriori.

The above boundary conditions, together with the choice of the seed configuration used in the numerical solver, select regular finite-energy configurations with a dipolar angular profile, characterized by a change of sign of the field across the equatorial plane, consistent with their $\ell=1$ nature. They are also consistent with the asymptotic behavior of the solutions at the boundaries of the integration domain. In particular, the small-$r$ expansion confirms regularity at the origin (see App.~\ref{app.expansion}). We construct these solutions numerically in Sec.~\ref{sec.numerical.rel.ell=1}.

%%%%%%%%%%%%%%%%%%%%%%%%%%%%%%%%%%%%%%%%%%%%%%%%%%%%%%
\subsection{Newtonian approximation}\label{Sec:newton}
%%%%%%%%%%%%%%%%%%%%%%%%%%%%%%%%%%%%%%%%%%%%%%%%%%%%%%

The Newtonian limit of the complex Einstein-Proca theory was derived in App.~B of Ref.~\cite{Nambo:2024hao}. Here we briefly summarize the main results, adapting the presentation to the notation and conventions used in this paper.

The Newtonian approximation arises from a combined nonrelativistic and weak-field expansion.\footnote{{The nonrelativistic expansion is an expansion of the matter sector in powers of the typical velocity, whereas the weak-field expansion is an expansion of the spacetime metric around Minkowski spacetime. Since these approximations act on different sectors of the theory, they are conceptually independent and neither implies the other in general. For self-gravitating virialized systems, however, the dynamics relates the corresponding expansion parameters through the virial theorem, $|\mathcal{U}|\sim v^2$, allowing both approximations to be consistently organized within the post-Newtonian expansion. Throughout this work, we retain only the leading-order terms in this combined expansion, corresponding to the Newtonian limit.}} In the nonrelativistic regime, the field frequency $\omega$ is close to the particle mass $\mu$, namely $\omega\to\mu$, and it is convenient to factor out the fast oscillatory phase associated with the rest-mass energy as\footnote{The relative factor of $1/\sqrt{2}$ with respect to Ref.~\cite{Nambo:2024hao} originates from the different normalization of the Einstein-Proca action; see footnote~\ref{footnote}.}
\begin{equation}
\mathcal{A} = \frac{1}{\sqrt{\mu}}
e^{-i\mu t}\left(a_0(t,\vec{x})\text{d}t +\psi_i(t,\vec{x}) \text{d}x^i\right).
\end{equation}
Here, $a_0$ and $\psi_i$ are slowly varying fields compared to the timescale $\mu^{-1}$.
In addition, in the weak-field limit the spacetime line-element can be expressed in the form
\begin{equation}\label{eq.metric.approx}
\text{d}s^2=-\left(1+2\Phi(t, \vec{x})\right)\text{d}t^2+\left(1-2\Psi(t, \vec{x})\right)\delta_{jk}\text{d}x^{j}\text{d}x^{k},
\end{equation}
where $\Phi$ and $\Psi$ are the gravitational potentials, assumed to be small. This expression is written in the Newtonian gauge, retaining only the scalar perturbations of the metric. Vector and tensor modes decouple at leading order in this limit and are therefore not relevant for our analysis.

The Newtonian limit is then implemented through the following power-counting scheme~\cite{Nambo:2024hao}:
\begin{equation}
\partial_t\sim\epsilon \mu,\quad \partial_i \sim \epsilon^{1/2}\mu,\quad a_0\sim \epsilon^{1/2}|\psi_i|, \quad \psi_i\sim \epsilon m_{\textrm{Pl}}\mu^{1/2}, \quad \Phi\sim\epsilon, \quad \Psi\sim \epsilon,
\end{equation}
where $\epsilon\ll 1$ is the small expansion parameter controlling the Newtonian regime and $m_{\textrm{Pl}}^{-2}=8\pi G$ is the reduced Planck mass. Under these assumptions, the Einstein-Proca action~(\ref{eq.EP.action}) reduces, at leading order in $\epsilon$, to
\begin{equation}\label{eq.Newtonian.action}
\mathcal{S}[\mathcal{U},\vec{\psi},\vec{\psi}^*]=\int dt\int dV \left[\frac{1}{8\pi G}\mathcal{U}\Delta \mathcal{U}-\mu \mathcal{U}n +\vec{\psi}^*\cdot\left(i\frac{\partial}{\partial t} +\frac{1}{2\mu}\Delta\right)\vec{\psi}\right].
\end{equation}
Here $\vec{\psi}$ is a complex-valued, three-component wave function encoding the spatial components of the vector field, $\mathcal{U}$ is the Newtonian gravitational potential (which, at this order, coincides with $\Phi=\Psi$), $dV$ and $\Delta$ refer to the volume element and Laplace operator, respectively, associated with three-dimensional Euclidean space, and $n = \vec{\psi}^{*}\cdot\vec{\psi}$ is the particle number density, which at leading order coincides with the Noether charge density $j^0$ introduced in Eq.~\eqref{eq.Q}. To obtain this expression, we have  used the Lorenz condition~\eqref{eq:lorenz}, which at leading order reduces to
\begin{equation}
a_0 = \frac{i}{\mu}\,\nabla\cdot\vec{\psi}.
\end{equation}
Thus, at this order, the temporal component $a_0$, being non-dynamical, is completely determined by the spatial wave function $\vec{\psi}$. It therefore plays the role of an auxiliary field that can be eliminated algebraically from the action, as we have done above, so that the dynamics is entirely encoded in the three spatial components of $\vec{\psi}$; see Eq.~(\ref{eq.Newtonian.action}).

In this reduced description, an important property of the Newtonian limit becomes manifest. As in the full Einstein-Proca theory, the action~(\ref{eq.Newtonian.action}) is invariant under global $U(1)$ phase transformations of the form $\vec{\psi}\mapsto e^{i\chi}\vec{\psi}$, with constant real $\chi$. However, in the nonrelativistic regime this symmetry is enhanced to $U(3)\supset U(1)$, and the action remains invariant under global transformations $\vec{\psi}\mapsto \hat{U}\vec{\psi}$, where $\hat{U}$ is a constant unitary $3\times3$ matrix. This enhancement is a distinctive feature of the Newtonian \mbox{spin-1} theory and has no counterpart in the relativistic Einstein-Proca system, whose symmetry is restricted to the global $U(1)$ phase invariance. Physically, this reflects the fact that, in the relativistic theory, the polarization degrees of freedom are tied to the spacetime structure through Lorentz symmetry and cannot be regarded as independent internal degrees of freedom. In the nonrelativistic limit, however, the three polarization states enter the leading-order action on an equal footing, and the theory acquires an additional internal $U(3)$ symmetry acting on the polarization space. The precise origin and realization of this  symmetry are discussed in App.~\ref{app.U3}.

\begin{figure}[t!]
\centering
\includegraphics[width=\textwidth]{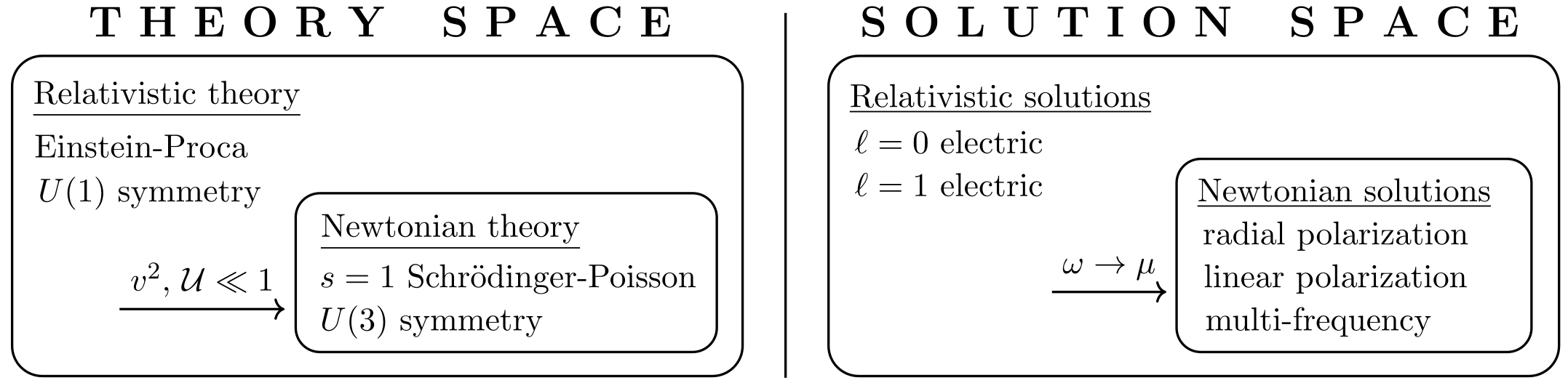}
\caption{\small\textbf{Relativistic versus Newtonian descriptions.} Schematic relation between the relativistic Einstein-Proca theory and its Newtonian limit, shown both at the level of the underlying theory and of its equilibrium solutions. From the theoretical perspective, the Newtonian limit corresponds to a weak-field, nonrelativistic expansion, under which the global symmetry is enhanced from $U(1)$ to $U(3)$. At the level of solutions, this limit is attained as $\omega\rightarrow\mu$, where relativistic Proca stars continuously reduce to their Newtonian counterparts: $\ell=0$ electric Proca stars map to radially polarized solutions, while $\ell=1$ electric Proca stars map to linearly polarized ones. The enlarged $U(3)$ symmetry gives rise to additional families of equilibrium solutions, including multi-frequency states that arise as extrema of the energy functional at fixed $\hat{Q}$.  }\label{fig.RelVSNew}
\end{figure}

The enlarged symmetry has important consequences for the structure of the solution space. Configurations that are physically distinct in the relativistic theory may become related by $U(3)$ transformations in the Newtonian limit and therefore belong to the same equivalence class from the nonrelativistic perspective. Conversely, the $U(3)$ invariance also supports additional families of static equilibrium configurations, such as the multi-frequency states, whose existence relies on the symmetry enhancement and for which no direct relativistic counterparts are presently known. As a result, the correspondence between relativistic and Newtonian solutions is considerably more subtle than in the scalar case. These features play a central role in the discussion below. Figure~\ref{fig.RelVSNew} provides a schematic illustration of this correspondence, both at the level of the theory and of its solution space.

Varying Eq.~(\ref{eq.Newtonian.action}) with respect to $\vec{\psi}^*$ and $\mathcal{U}$ yields
\begin{subequations}\label{eq.GPPs=1}
\begin{eqnarray}
& \displaystyle{i\frac{\partial\vec{\psi}}{\partial t} = -\frac{1}{2\mu}\Delta\vec{\psi}+\mu\mathcal{U}\vec{\psi},}\label{eq.GPs=1}\\
& \Delta\mathcal{U}  = 4\pi G \mu n, \label{eq.Poisson}
\end{eqnarray}
\end{subequations}
which constitute the \mbox{spin-1} Schr\"odinger-Poisson system. These equations describe the Newtonian limit of the Einstein-Proca theory and are expected to accurately capture its solutions whenever both the weak-field and nonrelativistic approximations apply. Moreover, they offer a considerably simpler framework for exploring this sector of the theory. The precise correspondence between relativistic and Newtonian solutions is, however, not entirely straightforward and will be discussed in Sec.~\ref{sec.relativistic_newtonian_conection}.

To facilitate this comparison, we  introduce the most relevant conserved quantities of the Newtonian system and discuss their relation to their relativistic counterparts. The action~(\ref{eq.Newtonian.action}) is invariant under global $U(3)$ transformations acting on $\vec{\psi}$. By Noether's theorem, this symmetry implies the conservation of the Hermitian rank-two tensor
\begin{equation}\label{eq.hat.Q}
\hat{Q}=\int (\vec{\psi}^{*}\otimes\vec{\psi}) dV.
\end{equation}
Among its associated conserved quantities, the most relevant for our purposes is the particle number,
\begin{equation}
N=\mathrm{Tr}(\hat{Q})=\int (\vec{\psi}^{*}\cdot\vec{\psi}) dV,
\end{equation}
which coincides with Eq.~(\ref{eq.Q}) in the nonrelativistic limit.\footnote{The spin angular momentum can also be extracted from $\hat{Q}$ according to $\vec{S}=-i\mathrm{Tr}(\hat{\varepsilon}\hat{Q})$, where $\hat{\varepsilon}$ denotes the Levi-Civita tensor and the trace represents the contraction of its last two indices with those of $\hat{Q}$. However, all configurations considered in this work have vanishing spin.}
The action~(\ref{eq.Newtonian.action}) is also invariant under time translations, $\vec{\psi}(t,\vec{x})\mapsto \vec{\psi}(t-t_0,\vec{x}),$ which implies the conservation of the Newtonian energy
\begin{equation}\label{eq.energy}
\mathcal{E}=\int\left(\frac{1}{2\mu}|\nabla\vec{\psi}|^2+\frac{\mu}{2}n\mathcal{U}\right)dV.
\end{equation}
This quantity represents the nonrelativistic contribution to the total energy and consists of kinetic and gravitational terms. To relate the ADM mass $M$ introduced in the relativistic theory with the Newtonian energy $\mathcal{E}$, recall that, in the Newtonian limit, $M=\mu N+\mathcal{E}$, where $\mu N$ is the rest-mass contribution. Comparing this expression with the definition of the binding energy in Eq.~(\ref{eq.binding.energy}), and recalling that the configurations considered here belong to the particle sector, for which $N>0$, we obtain
\begin{equation}
E_{\rm bind}=M-\mu N=\mathcal{E}.
\end{equation}
Thus, the conserved Newtonian energy $\mathcal{E}$ is precisely the leading-order contribution to the relativistic binding energy in the Newtonian limit.

%%%%%%%%%%%%%%%%%%%%%%%%%%%%%%%%%%%%%%%%%%%%%%%%%%%
\subsection{Newtonian solitons and boundary conditions}\label{sec.Newtonian. ansatz}
%%%%%%%%%%%%%%%%%%%%%%%%%%%%%%%%%%%%%%%%%%%%%%%%%%%

We now turn to  the study of solitonic solutions of the \mbox{spin-1} Schr\"odinger-Poisson system described by Eqs.~\eqref{eq.GPPs=1}. For the purposes of the present work, it is sufficient to consider stationary and spherically symmetric configurations, since, although this is not immediately apparent, these already capture the solutions relevant for comparison with their relativistic counterparts presented in Sec.~\ref{sec:2.2}. The full spectrum of spherical equilibrium configurations, including both stationary and multi-frequency states, was presented in Ref.~\cite{Nambo:2024hao}.

For stationary configurations, the Newtonian gravitational potential is time independent, while the vector wave function exhibits harmonic time dependence,
\begin{subequations}\label{eq.New.stationary}
\begin{equation}
\mathcal{U}(t,\vec{x}) = \mathcal{U}(\vec{x}),\quad \vec{\psi}(t,\vec{x}) = e^{-iEt}\vec{\sigma}(\vec{x}).
\end{equation}
Here $E$ is a real frequency parameter, which is negative for the localized bound states considered in this paper, and $\vec{\sigma}$ captures the spatial profile of the vector field. Without loss of generality, this spatial profile can be decomposed as
\begin{equation}
\vec{\sigma}(\vec{x})=\sigma(\vec{x})\hat{\epsilon}(\vec{x}),
\end{equation}
\end{subequations}
where $\sigma$ is a real scalar function of the spatial coordinates and $\hat{\epsilon}$ is a unit polarization vector, which may in general depend also on the spatial coordinates and satisfies $\hat{\epsilon}^*\cdot\hat{\epsilon}=1$. This decomposition is particularly useful because it separates the density profile of the configuration, as determined by the particle number density $n=\vec{\psi}^*\cdot\vec{\psi}$,  from its local polarization structure. It is worth emphasizing that, as in the relativistic case, although the wave function $\vec{\psi}$ carries the harmonic phase factor $e^{-iEt}$, physically observable quantities such as the particle number density and the gravitational potential are time independent.

We further restrict our attention to spherically symmetric configurations, defined as those
invariant under spatial rotations. The Newtonian action, Eq.~(\ref{eq.Newtonian.action}), is invariant under both of the following representations of the $SO(3)$ group:
\begin{equation}\label{eq.SO3reps}
\vec{\psi}(t,\vec{x})\mapsto \hat{R}\vec{\psi}(t,\hat{R}^{-1}\vec{x}),\quad \vec{\psi}(t,\vec{x})\mapsto \vec{\psi}(t,\hat{R}^{-1}\vec{x}),
\end{equation}
where $\hat{R}$ denotes a rotation matrix.  Accordingly, a spherically symmetric configuration is one that transforms trivially under either of these actions of the rotation group. Under this restriction, the vector field factorizes into a fixed polarization structure and a scalar radial profile, reducing the problem effectively to a single radial function $\sigma(r)$. This naturally leads to the two representative classes of solutions that we analyze below.\footnote{In addition to radially and linearly polarized states, the Newtonian theory also admits all configurations related to them by global unitary transformations, such as helicoidally or circularly polarized states. Owing to the $U(3)$ symmetry of the Newtonian theory, all such configurations are physically equivalent in the sense that they share the same density profile, gravitational potential, and total energy, differing only in their polarization structure. For this reason, we do not discuss them separately. This degeneracy is lifted in the full relativistic limit, where the symmetry is reduced to the global $U(1)$ phase invariance of the Einstein-Proca theory. We return to this point later.}

%%%%%%%%%%%%%%%%%%%%%%%%%%%%%%%%%%%%%%%%%%%%%%%%
\subsubsection{Radially polarized Proca stars}
%%%%%%%%%%%%%%%%%%%%%%%%%%%%%%%%%%%%%%%%%%%%%%%%

Radially polarized Proca stars are characterized by a polarization vector that points along the radial direction~\cite{Nambo:2024hao}, 
\begin{equation}
\hat{\epsilon} = \hat{e}_r.
\end{equation}
These configurations are also commonly referred to as hedgehog configurations, reflecting the fact that the polarization at each point is oriented away from the soliton center. In this case, the physically observable quantities, such as the particle number density and the gravitational potential, are manifestly spherically symmetric as a direct consequence of the radial structure of the polarization vector.

The boundary conditions are determined by imposing regularity at the origin and localization at spatial infinity. In particular, regularity at $r=0$ requires the full vector field $\vec{\psi}$ to remain smooth and finite. Since the unit vector $\hat{e}_r$ is not well defined at the origin, this implies that the radial amplitude must vanish at least linearly as $r\to0$. Accordingly, we impose
\begin{subequations}\label{eqs.boundary.radial}
\begin{equation}
\left.\partial_r\mathcal{U}(r)\right|_{r=0}=0,\quad \left.\sigma(r)\right|_{r=0}=0,
\end{equation}
with $\left.\partial_r\sigma(r)\right|_{r=0}$ finite. At spatial infinity, matter fields must decay sufficiently rapidly to ensure a finite total mass and particle number, namely
\begin{equation}
\lim_{r\to\infty}\mathcal{U}(r)=0,\quad \lim_{r\to\infty}\sigma(r)=0.
\end{equation}
\end{subequations}

These boundary conditions define regular, self-gravitating radially polarized configurations. Their numerical construction is presented in Sec.~\ref{sec.numerical.newtonian.radial}.

%%%%%%%%%%%%%%%%%%%%%%%%%%%%%%%%%%%%%%%%%%%%%%%%%
\subsubsection{Linearly polarized Proca stars}
%%%%%%%%%%%%%%%%%%%%%%%%%%%%%%%%%%%%%%%%%%%%%%%%%%

We next consider linearly polarized Proca stars, which are characterized by a constant polarization vector~\cite{Nambo:2024hao}. Without loss of generality, this can be chosen to point along the $z$-direction: 
\begin{equation}
\hat{\epsilon} = \hat{e}_z.
\end{equation}
With this choice, the vector field is everywhere aligned along a fixed spatial direction, while its amplitude depends only on the radial coordinate. As a result, the particle number density and the gravitational potential remain spherically symmetric, even though the vector field itself selects a preferred direction in space.\footnote{Although these configurations may not appear spherically symmetric from their polarization structure alone, they nevertheless remain invariant under spatial rotations because the full configuration transforms according to the trivial representation under the $SO(3)$ action defined by the second expression in Eq.~(\ref{eq.SO3reps}). See Sec.~IV of Ref.~\cite{Nambo:2024hao} for a more detailed discussion.}

As in the radial case, the corresponding physical solutions are determined by regularity at the origin and localization at spatial infinity. Here, regularity at $r=0$ requires both the gravitational potential and the radial field profile to be smooth and free of singularities, leading to
\begin{subequations}\label{eqs.boundary.linear}
\begin{equation}
\left.\partial_r\mathcal{U}(r)\right|_{r=0}=0,\quad \left.\partial_r\sigma(r)\right|_{r=0}=0.
\end{equation}
At spatial infinity, the matter fields must decay sufficiently rapidly so that the configuration remains localized and possesses finite total mass and particle number:
\begin{equation}
\lim_{r\to\infty}\mathcal{U}(r)=0,\quad \lim_{r\to\infty}\sigma(r)=0.
\end{equation}
\end{subequations}

These boundary conditions define regular, self-gravitating solitonic configurations of constant polarization. The numerical construction of these solutions is presented in Sec.~\ref{sec.numerical.newtonian.linear}.

%%%%%%%%%%%%%%%%%%%%%%%%%%%%%%%%%%%%%%%%%%%%%%%%%%%%
\section{Numerical solutions}\label{sec.numerical}
%%%%%%%%%%%%%%%%%%%%%%%%%%%%%%%%%%%%%%%%%%%%%%%%%%%%

To the best of our knowledge, no analytical Proca star solutions are known, either in the full relativistic theory or in its Newtonian limit, and the corresponding Einstein-Proca and \mbox{spin-1} Schr\"odinger-Poisson equations must therefore be solved numerically. For the class of Proca star ans\"atze considered in this work, such numerical constructions have been carried out previously in, e.g., Refs.~\cite{Brito:2015pxa,{Herdeiro:2023wqf},Nambo:2024hao}. Nevertheless, we revisit the numerical analysis here for two main reasons. First, it provides the necessary dataset for the comparison between relativistic and Newtonian solutions presented in Sec.~\ref{sec.relativistic_newtonian_conection}. Second, it allows us to present the results in a unified framework and clarify several subtle points that are important for their interpretation.

%%%%%%%%%%%%%%%%%%%%%%%%%%%%%%%%%%%%%%%%%%%%%%%%%%%%%%%%
\subsection{Relativistic Proca stars}\label{sec.numerical.rel}
%%%%%%%%%%%%%%%%%%%%%%%%%%%%%%%%%%%%%%%%%%%%%%%%%%%%%%%%

To solve the Einstein-Proca equations numerically, it is convenient to work with dimensionless variables defined by
\begin{equation}\label{eq.dimensionless.relativistic}
t_{\textrm{phys}}:=\frac{1}{\mu}t_{\textrm{R}},\quad 
\vec{x}_{\textrm{phys}}:=\frac{1}{\mu}\vec{x}_{\textrm{R}},\quad 
\mathcal{A}^{\mu}_{\textrm{phys}} := \frac{1}{\sqrt{4\pi G}}\mathcal{A}^{\mu}_{\textrm{R}},
\end{equation}
where the subscript {\it R} denotes dimensionless quantities used in the relativistic code. Since the frequency has dimensions of inverse time, it rescales accordingly as $\omega_{\textrm{phys}}:=\mu \omega_{\textrm{R}}$. By contrast, the metric components are already dimensionless and therefore require no rescaling.
This choice absorbs the relevant mass and gravitational scales, leaving a system that depends only on dimensionless variables. For notational simplicity, throughout this section we suppress the subscript {\it R} and work exclusively with dimensionless quantities, unless otherwise stated. In Sec.~\ref{sec.relativistic_newtonian_conection}, where both relativistic and Newtonian variables are used, we will reinstate this notation to avoid ambiguity.

Substituting the static axisymmetric metric ansatz~\eqref{eq.metric}, the electric Proca ansatz~\eqref{eq.ansatz.electric}, and the dimensionless code variables introduced in Eq.~\eqref{eq.dimensionless.relativistic} into the field equations~\eqref{eq.EP} yields a system of coupled partial differential equations for the metric components and the Proca field amplitudes. Due to their length, we do not present them explicitly here and instead refer the reader to App.~\ref{app.eqs}.
To obtain physically acceptable solutions, this system must be supplemented with appropriate boundary conditions. These were introduced in Sec.~\ref{sec:2.2}, and their explicit form depends on the multipolar sector under consideration. 

In what follows, we consider the $\ell=0$ and $\ell=1$ electric cases, for which the numerical strategies differ. For the monopolar $\ell=0$ configurations, spherical symmetry reduces the field equations to a system of ordinary differential equations. We solve the resulting boundary-value problem using a shooting method combined with a Runge-Kutta integrator, following Ref.~\cite{Brito:2015pxa}. For the dipolar $\ell=1$ configurations, by contrast, axial symmetry still leaves a genuinely two-dimensional system of coupled elliptic partial differential equations. We therefore solve the equations as a two-dimensional boundary-value problem using the CADSOL package~\cite{Schonauer:1989,Schauder:1992}, which employs finite-difference discretization and Newton-Raphson iteration, following Ref.~\cite{Herdeiro:2023wqf, Gervalle:2026let}. We compactify the radial coordinate according to $x=r/(1+r)$, so that the computational domain is $0\leq x\leq1$, and restrict the angular domain to $0\leq\theta\leq\pi/2$ by exploiting the reflection symmetry of the solutions. A typical grid consists of $400\times60$ points in the radial and angular directions, respectively. We generate sequences of dipolar configurations by varying the frequency parameter $\omega$, using each converged solution as the initial guess for the next one. CADSOL provides several error estimates that allow us to assess the accuracy of the numerical solutions. The typical numerical error for the dipolar solutions is of order of $10^{-5}$. Further details of the numerical implementation can be found in the cited references.

%%%%%%%%%%%%%%%%%%%%%%%%%%%%%%%%%%%%%%%%%%%%%%%%%%%%%%%%%%%%%%%%%%%%%%%%%%%%%%
\subsubsection{\texorpdfstring{$\ell=0$}{l=0} electric Proca stars}\label{sec.numerical.rel.ell=0} 
%%%%%%%%%%%%%%%%%%%%%%%%%%%%%%%%%%%%%%%%%%%%%%%%%%%%%%%%%%%%%%%%%%%%%%%%%%%%%%

\begin{figure}[t!]
\centering
\includegraphics[width=\textwidth]{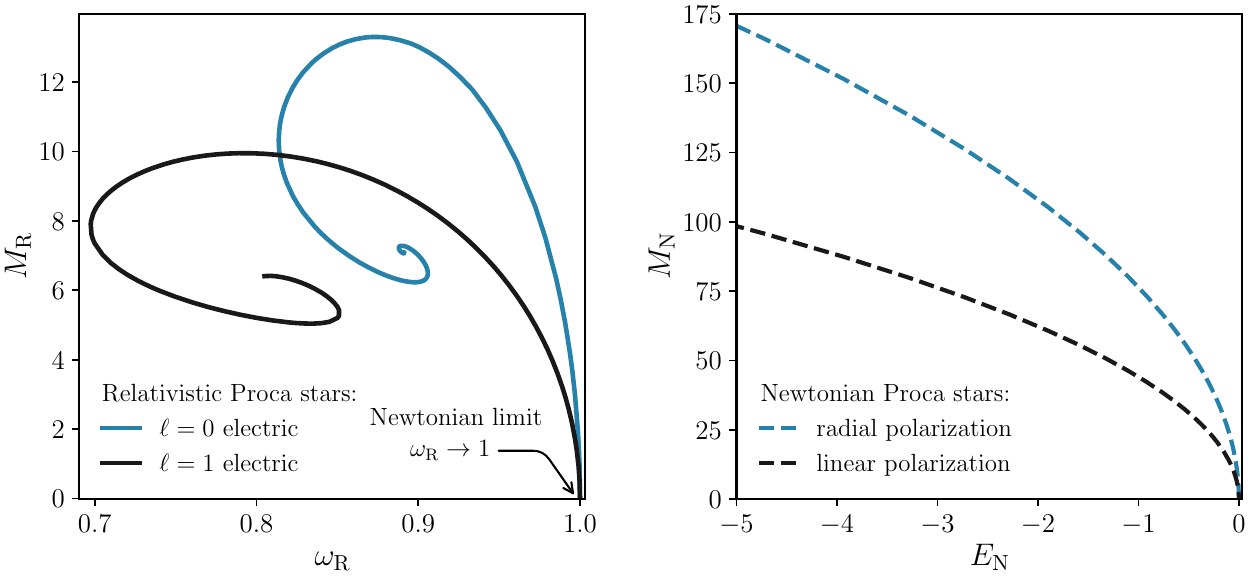}
\caption{\small\textbf{Mass-frequency diagram.} Mass as a function of frequency along the fundamental branch ($n=0$) for representative solutions in the different regimes of the theory. {\it Left panel:} Relativistic monopolar ($\ell=0$) and dipolar ($\ell=1$) electric Proca star solutions. {\it Right panel:} Newtonian radially and linearly polarized Proca star solutions. All quantities are shown in the dimensionless code variables defined in the main text. The lower-right region of the left panel, marked by an arrow, indicates where the Newtonian solutions reside.}\label{fig:pspace}
\end{figure}

Equations~(\ref{eqs.system.electric}), together with the boundary conditions introduced in~\eqref{eq.boundary.ell=0.electric}, define a nonlinear eigenvalue problem for the frequency $\omega$~\cite{Brito:2015pxa}. For each choice of the central amplitude of the field $V$, namely $V_0 \equiv V(r=0)$, there exists a discrete set of eigenfrequencies $\omega_n(V_0)$, with $n = 0,1,2,\ldots$, for which the boundary conditions are satisfied.\footnote{The values of the metric functions at the origin can be chosen arbitrarily, since constant shifts $F_i(r)\mapsto F_i(r)+C_i$ can be absorbed by coordinate rescalings. In particular, a constant shift in $F_0(r)$ corresponds to a rescaling of the time coordinate, while a constant shift in $F_1(r)$ can be absorbed into a rescaling of the radial coordinate. One may therefore fix convenient values for these functions at the origin during the numerical integration and subsequently use the corresponding coordinate rescalings to impose the asymptotic conditions $\lim_{r\to\infty}F_i(r)=0$, thereby bringing the asymptotic metric into its standard Minkowski form. In addition, the value of $\partial_r H_1(r)$ at the origin is not an independent parameter, but is determined by the central amplitude $V_0$ through the Lorenz condition.} 

These solutions can be organized into families labeled by the integer $n$, which counts the number of nodes of the function $H_1(r)$ in the interval $0<r<\infty$.\footnote{For the monopolar case, the number of nodes of the function $V(r)$ exceeds that of $H_1(r)$ by one. This can be appreciated, for instance, in Figs.~\ref{Fig:Spherical0.95} and~\ref{Fig:Spherical0.999}.}
As $V_0$ is varied, each family traces out a continuous branch in parameter space, typically exhibiting a maximum ADM mass together with a turning point in the frequency. The fundamental branch ($n=0$) is shown by the blue curve in the left panel of Fig.~\ref{fig:pspace}, where each point corresponds to a different value of the central amplitude $V_0$.
Traditionally, the portion of this branch lying to the right of the maximum mass configuration has been identified as stable against radial perturbations, while the left branch is unstable~\cite{Brito:2015pxa}. However, more recent analyses~\cite{Herdeiro:2023wqf} have shown that the spherical fundamental branch is unstable under generic non-spherical perturbations and should therefore not be regarded as the true ground state of the relativistic theory (see below for further discussion). Instead, it behaves as an excited configuration that can dynamically evolve toward a lower energy nonspherical branch. For $n>0$, all solutions are found to be unstable.

\begin{figure}[t!]
\centering
\includegraphics[width=\textwidth]{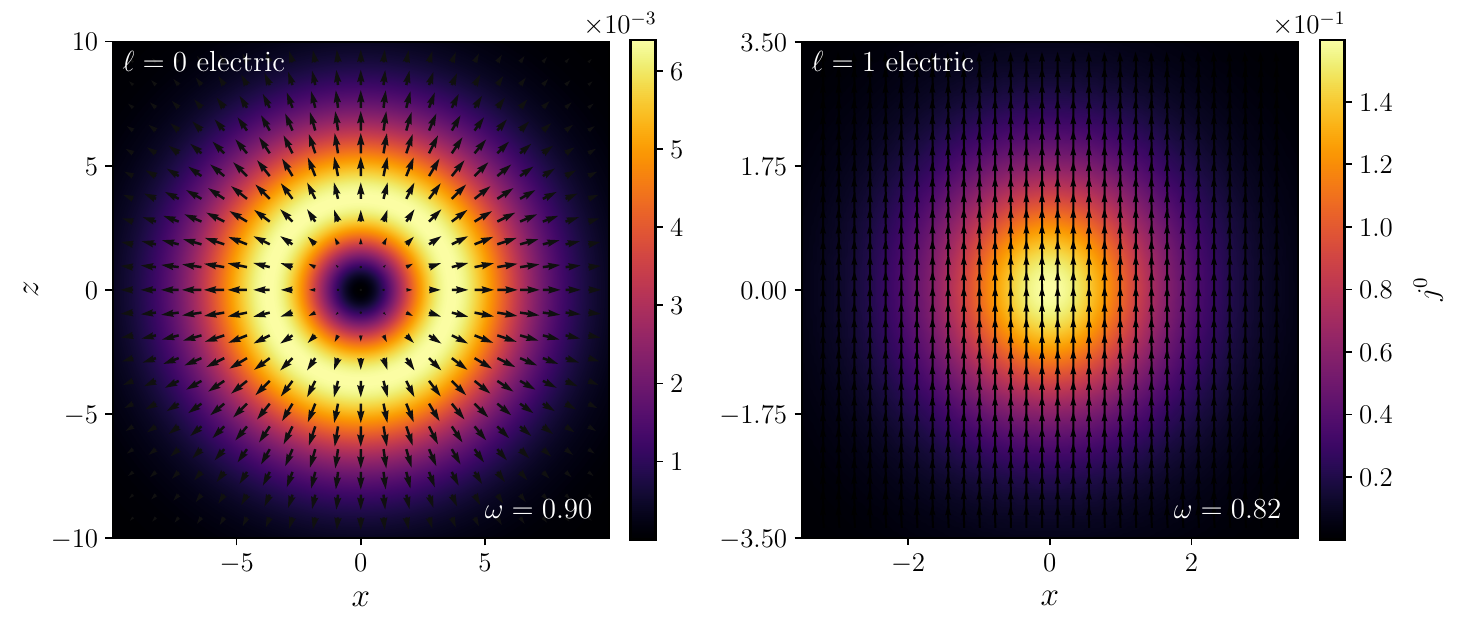}
\caption{\small\textbf{Relativistic Proca stars.}  Real part of the spatial components of the Proca field, $\textrm{Re}(\mathcal{A}_i)$ [arrows], and  particle number density, $j^0$ [color shading], for representative relativistic Proca star solutions at $t=0$. {\it Left panel:} Monopolar ($\ell = 0$) electric Proca star. {\it Right panel:} Dipolar ($\ell= 1$) electric Proca star. The frequencies were chosen to be $\omega = 0.90$ and $\omega = 0.82$, respectively, placing both solutions well within the relativistic regime and close to their maximum ADM mass configuration.  }\label{fig.relativisticPS}
\end{figure}

For reference, the configuration with $\omega=0.900$ is shown in Fig.~\ref{fig.relativisticPS}, left panel. As expected for the $\ell=0$ sector, the corresponding solution is manifestly spherically symmetric and features a central hole in the particle number density.

%%%%%%%%%%%%%%%%%%%%%%%%%%%%%%%%%%%%%%%%%%%%%%%%%%%%%%%%%%%%%%%%%%%%%%%%%%%%%%%%
%\subsubsection{$\ell=1$ electric Proca stars}\label{sec.numerical.rel.ell=1} 
\subsubsection{\texorpdfstring{$\ell=1$}{l=1} electric Proca stars}\label{sec.numerical.rel.ell=1} 
%%%%%%%%%%%%%%%%%%%%%%%%%%%%%%%%%%%%%%%%%%%%%%%%%%%%%%%%%%%%%%%%%%%%%%%%%%%%%%%%

\begin{figure}[t!]
\centering
\includegraphics[width=\textwidth]{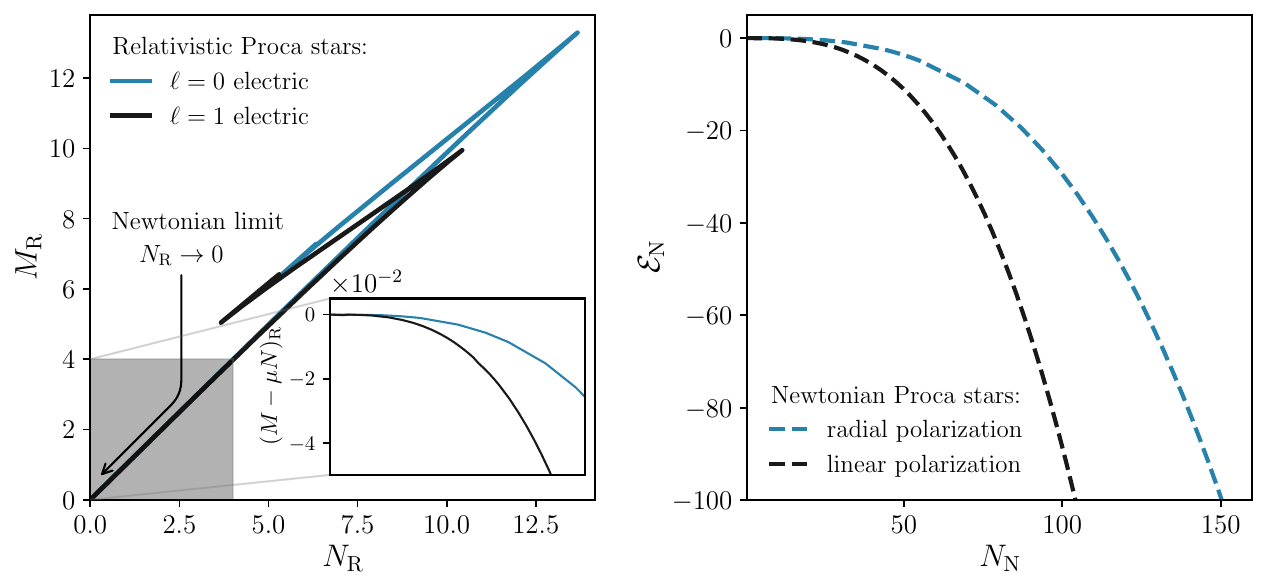}
\caption{\small\textbf{Energy-particle number diagram.} Energy as a function of particle number along the fundamental branch ($n=0$) for representative solutions in the different regimes of the theory. {\it Left panel:} Relativistic monopolar ($\ell=0$) and dipolar ($\ell=1$) electric Proca star solutions. {\it Right panel:} Newtonian radially and linearly polarized Proca star solutions. The inset in the left panel shows the binding energy, facilitating comparison with the Newtonian energy displayed in the right panel. In the relativistic theory, all dynamically stable solutions belong to the $\ell=1$ branch and, at fixed particle number, have the lowest energy, suggesting that they may represent the ground state. By contrast, in the Newtonian limit, both families are stable.}\label{fig:NvsE}
\end{figure}

If, instead of imposing the boundary conditions~\eqref{eq.boundary.ell=0.electric}, we adopt those appropriate to the dipolar ($\ell=1$) electric ansatz, given in Eqs.~\eqref{eq.boundary.ell=1.electric}, the solutions of the system~\eqref{eqs.system.electric} differ qualitatively from those discussed in the previous subsection.

As in the $\ell = 0$ case, Eqs.~\eqref{eqs.system.electric}, together with the boundary conditions~\eqref{eq.boundary.ell=1.electric}, define a nonlinear eigenvalue problem for the frequency $\omega$~\cite{Herdeiro:2023wqf}. However, regularity at the origin now requires $\left.V(r,\theta)\right|_{r=0}=0$, and the central value of this function can no longer be used to parametrize the solutions. Instead, one must consider the leading behavior in a near-origin expansion, which for the dipolar sector takes the form $V(r,\theta)= V_1 r\cos\theta +\ldots$, and use the coefficient $V_1$ to label the different $\ell=1$ solutions (see App.~\ref{app.expansion}). In particular, and similar to the $\ell= 0$ case, each value of $V_1$ admits a discrete set of eigenfrequencies $\omega_n(V_1)$, with $n = 0,1,2,\ldots$, for which the boundary conditions are satisfied.

As in the monopolar case, these solutions can be organized into families labeled by the integer $n$, which counts the number of nodes of the function $H_1(r)$ in the interval $0<r<\infty$.\footnote{For the dipolar case, the number of nodes of $H_1(r)$ coincides with that of $V(r)$ and $H_2(r)$. This can be appreciated, for instance, in Figs.~\ref{Fig:Prolate0.999n=0} and~\ref{Fig:Prolate0.999n=1}.} Varying the amplitude $V_1$ generates a continuous branch in parameter space, which again exhibits a maximum ADM mass and a corresponding turning point in the frequency. The fundamental family ($n=0$) is represented by the black curve in the left panel of Fig.~\ref{fig:pspace}. In contrast to the $\ell=0$ family, however, the portion of the $\ell=1$ branch lying to the right of the maximum mass configuration has been found to be dynamically stable under generic, including non-spherical, perturbations, while the branch to the left remains unstable~\cite{Herdeiro:2023wqf}.

For reference, the configuration with $\omega=0.820$ is shown in the right panel of Fig.~\ref{fig.relativisticPS}. As expected for the $\ell=1$ sector, the corresponding solution is not spherically symmetric. In particular, the isosurfaces of constant particle number are prolate, and the vector field exhibits the characteristic dipolar angular dependence of this multipole. This is more clearly illustrated in the first row of Fig.~\ref{fig.eccentricity}, where the contours of constant particle number highlight the nonspherical shape of the configuration.

\subsubsection{Energy considerations}

To further clarify the energetic properties of these solutions, in the left panel of Fig.~\ref{fig:NvsE} we show the ADM mass $M$ as a function of the particle number for the $\ell=0$ and $\ell=1$ electric Proca stars along the fundamental family. Since the spacetimes under consideration are asymptotically flat, the ADM mass provides the appropriate notion of total energy for these configurations. A notable feature of this diagram is that, at fixed particle number, the dynamically stable $\ell=1$ configurations consistently possess lower energy than both the unstable $\ell=1$ solutions that coexist at the same particle number and the $\ell=0$ configurations, which are unstable throughout the family. Moreover, excited families ($n>0$) and higher-multipole solutions ($\ell>1$), which are not shown in the figure, lie at even higher energy and are likewise unstable. 

Taken together, this energetic ordering and the corresponding dynamical stability suggest that the stable $\ell=1$, $n=0$ configurations may represent the ground-state family of the complex Einstein-Proca theory~\cite{Herdeiro:2023wqf}. Nevertheless, this conclusion has not been established rigorously, and the existing evidence is primarily indirect. The positive-energy theorems~\cite{Schoen1979OnTP,Schoen1981ProofOT,Witten1981ANP} guarantee, under their respective assumptions, that the ADM mass of an asymptotically flat spacetime satisfying the dominant energy condition is non-negative, $M\ge 0$. This establishes a lower bound on the total energy, but does not imply that, for a given nonzero particle number $N$, the energy attains a minimum within the space of nontrivial Einstein-Proca configurations. In particular, these theorems neither identify a minimizing family at fixed $N$ nor establish that such a constrained minimum exists. To the best of our knowledge, no rigorous variational result establishing the existence and identifying the minimizer of the energy at fixed particle number is currently available for the relativistic theory under consideration. Thus, while the stable $\ell=1$, $n=0$ family may be regarded as a {\it candidate ground-state family}, the existence of a ground state in the precise variational sense defined above has not been established for the relativistic theory. This contrasts with the Newtonian limit, where the analogous constrained minimization problem can be established mathematically~\cite{Nambo:2024hao}. We return to this point in Sec.~\ref{sec.numerical.newtonian}, when discussing the solutions of the \mbox{spin-1} Schr\"odinger-Poisson equations, and later in Sec.~\ref{sec.relativistic_newtonian_conection}, when comparing these solutions with their Newtonian counterparts.

%%%%%%%%%%%%%%%%%%%%%%%%%%%%%%%%%%%%%%%%%%%%%%%%%%%%%%%%%%%%%%%%%%%
\subsection{Newtonian Proca stars}\label{sec.numerical.newtonian}
%%%%%%%%%%%%%%%%%%%%%%%%%%%%%%%%%%%%%%%%%%%%%%%%%%%%%%%%%%%%%%%%%%%

As in the relativistic case, it is convenient to introduce dimensionless variables in order to solve the \mbox{spin-1} Schr\"odinger-Poisson  equations numerically. These are defined as
\begin{subequations}\label{eq.dimensionless.Newtonian}
\begin{equation}
t_{\textrm{phys}}:=\frac{\lambda_*}{4\pi G\mu^3}t_{\textrm{N}},\quad \vec{x}_{\textrm{phys}}:=\frac{\lambda_*^{1/2}}{\sqrt{8\pi G}\mu^2}\vec{x}_{\textrm{N}},\quad \mathcal{U}_{\textrm{phys}}:= \frac{4\pi G \mu^2}{\lambda_*}\mathcal{U}_{\textrm{N}},\quad 
\vec{\psi}_{\textrm{phys}} := \frac{\sqrt{8\pi G}\mu^{5/2}}{\lambda_*}\vec{\psi}_{\textrm{N}},
\end{equation}
where the subscript {\it N} denotes dimensionless quantities used in the Newtonian code. The frequency scales with the inverse of the time unit and therefore transforms as
\begin{equation}
E_{\textrm{phys}}:=\frac{4\pi G\mu^3}{\lambda_*}E_{\textrm{N}}.
\end{equation}
\end{subequations}
As before, throughout the present section we suppress the subscript {\it N} for notational simplicity and reintroduce it in Sec.~\ref{sec.relativistic_newtonian_conection}, where both relativistic and Newtonian variables are used simultaneously. 

Here $\lambda_*$ is an arbitrary positive constant that reflects the scaling symmetry of the \mbox{spin-1} Schr\"odinger-Poisson system~(\ref{eq.GPPs=1}),
\begin{equation}\label{eq.scaling.free}
t\mapsto\lambda_*^{-1}t, \quad 
\vec{x}\mapsto\lambda_*^{-1/2}\vec{x}, \quad 
\mathcal{U}\mapsto\lambda_* \mathcal{U},\quad \vec{\psi} \mapsto \lambda_*\vec{\psi}.
\end{equation}
This symmetry implies that the Newtonian equations admit continuous families of  solutions related by a simple rescaling. This freedom will play an important role in the next section, where the Newtonian and relativistic solutions are compared. In practice, in Eq.~\eqref{eq.dimensionless.Newtonian}, the constant $\lambda_*$ may be fixed by normalizing a convenient quantity, such as the central field amplitude or the total mass, or, alternatively, one may simply set $\lambda_*=1$ without loss of generality.

Substituting the stationary and spherically symmetric ansatz~\eqref{eq.New.stationary}, with $\sigma(\vec{x})=\sigma(r)$ and $\hat{\epsilon}(\vec{x})=\hat{e}_r$ or $\hat{e}_z$ depending on the polarization, into the field equations~\eqref{eq.GPPs=1}, one obtains a system of coupled ordinary differential equations for the radial profile and the gravitational potential. Expressed in terms of the dimensionless variables introduced in Eq.~\eqref{eq.dimensionless.Newtonian}, this system takes the form:
\begin{subequations}\label{eqs.system.Newtonian}
\begin{eqnarray}
&\displaystyle{\left(\Delta_s -\frac{2\gamma}{r^2} \right)\sigma +\mathcal{U}\sigma =E\sigma,} \label{eqs.system.Newtonian.1}\\
&\Delta_s \mathcal{U} = \sigma^{2},
\end{eqnarray}
\end{subequations}
where $\Delta_s:=\frac{1}{r}\frac{d^2}{dr^2}r$ denotes the radial part of Laplace operator, and
the parameter $\gamma$ distinguishes the polarization structure of the solution, taking the values $\gamma=1$ for radial polarization and $\gamma=0$ for linear polarization. 

To obtain physically acceptable solutions, this system must be supplemented with appropriate boundary conditions, whose precise form depends on the polarization sector under consideration.  The numerical solutions are obtained using an adaptive explicit Runge-Kutta 5(4) integrator combined with a shooting method based on bisection. Since the numerical integration reaches only a finite radius, the solutions are matched to their analytic asymptotic behavior at large radii. Further details of the numerical implementation can be found in Refs.~\cite{Roque:2023sjl, Nambo:2025lnu}. We next discuss the radial and linear cases separately.

%To obtain physically acceptable solutions, this system must be supplemented with appropriate boundary conditions, whose precise form depends on the polarization sector under consideration. We next discuss the radial and linear cases separately.

%%%%%%%%%%%%%%%%%%%%%%%%%%%%%%%%%%%%%%%%%%
\subsubsection{Radially polarized Proca stars}\label{sec.numerical.newtonian.radial}
%%%%%%%%%%%%%%%%%%%%%%%%%%%%%%%%%%%%%%%%%

\begin{figure}[t!]
\centering
\includegraphics[width=\textwidth]{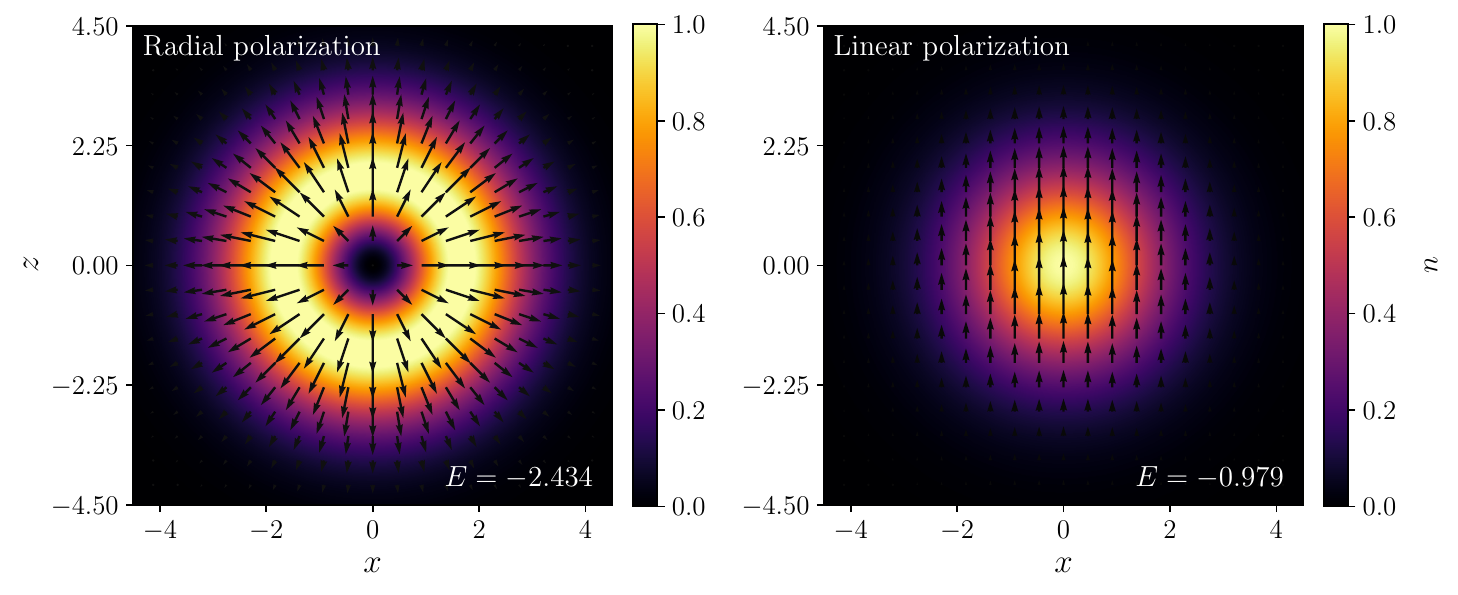}
\caption{\small\textbf{Newtonian Proca stars.} Real part of the vector wave function, $\textrm{Re}(\vec{\psi})$ [arrows], and particle number density, $n$ [color shading], for representative Newtonian Proca star solutions at $t=0$. {\it Left panel:} Radially polarized Proca star with central slope $\sigma_0'=1$. {\it Right panel:} Linearly polarized Proca star with central amplitude $\sigma_0=1$. The corresponding energy eigenvalues are $E = - 2.434$ and $E = - 0.979$, respectively.}\label{fig.NewtonianPS}
\end{figure}

For radially polarized configurations, the polarization-dependent parameter in Eqs.~(\ref{eqs.system.Newtonian}) takes the fixed value $\gamma=1$. Together with the boundary conditions introduced in~\eqref{eqs.boundary.radial}, this defines a nonlinear eigenvalue problem for the frequency $E$~\cite{Nambo:2024hao}. For each choice of the central slope of the vector field, $\sigma_0'\equiv\partial_r\sigma(r)|_{r=0}$, there exists a discrete set of frequencies $E_n(\sigma_0')$, with $n=0,1,2,\ldots$, for which the boundary conditions are satisfied.\footnote{The \mbox{spin-1} Schr\"odinger-Poisson equations are invariant under simultaneous shifts of the frequency and the gravitational potential, namely $E\mapsto E+C$ and $\mathcal{U}(r)\mapsto \mathcal{U}(r)+C$, with $C$ constant. As a result, only potential differences are physically meaningful. One may therefore fix $\mathcal{U}(r=0)$ arbitrarily during the numerical integration (for instance, to unity) and subsequently shift the solution so that $\lim_{r\to\infty}\mathcal{U}(r)=0$, as is customary.}

The solutions are naturally organized into families labeled by the integer $n$, which counts the number of radial nodes of $\sigma(r)$ in the interval $0<r<\infty$. As $\sigma_0'$ is varied, each family traces out a continuous branch in parameter space. The fundamental branch ($n=0$) is shown by the blue curve in the right panel of Fig.~\ref{fig:pspace}, where each point corresponds to a different value of $\sigma_0'$.
Owing to the scaling symmetry~\eqref{eq.scaling.free}, all solutions along a given branch are related by a global rescaling and therefore share the same dimensionless profile up to an overall choice of scale. In contrast to the relativistic case, there is no maximum-mass configuration, and instead the mass increases monotonically as the frequency decreases toward more negative values. Moreover, all configurations along the fundamental branch have been shown to be linearly stable under arbitrary perturbations, whereas the excited branches ($n>0$), which are not shown in the figure, are unstable~\cite{Nambo:2025lnu}.

For reference, the configuration with $\sigma_0'=1$ is shown in Fig.~\ref{fig.NewtonianPS}, left panel. As expected for the radial sector, the corresponding solution is manifestly spherically symmetric, both in the particle number density and in its polarization structure.

%%%%%%%%%%%%%%%%%%%%%%%%%%%%%%%%%%%%%%%%%%%%%%%%
\subsubsection{Linearly polarized Proca stars}\label{sec.numerical.newtonian.linear}
%%%%%%%%%%%%%%%%%%%%%%%%%%%%%%%%%%%%%%%%%%%%%%%

The analysis for linearly polarized configurations proceeds in a completely analogous way, the only difference being that the polarization parameter now takes the value $\gamma=0$, and the appropriate boundary conditions are those in Eqs.~\eqref{eqs.boundary.linear}. In this case, for each choice of the central amplitude, $\sigma_0\equiv \sigma(r=0)$, one obtains a discrete set of frequencies $E_n(\sigma_0)$, organized into families labeled by the node number $n$.

The fundamental branch ($n=0$) is shown by the black curve in the right panel of Fig.~\ref{fig:pspace}. As in the radial case, all solutions along this branch are related by the scaling symmetry~\eqref{eq.scaling.free}, the mass increases monotonically as the frequency decreases, and the entire fundamental branch is linearly stable, while the excited branches ($n>0$) are unstable~\cite{Nambo:2025lnu}.

For reference, the configuration with $\sigma_0=1$ is shown in Fig.~\ref{fig.NewtonianPS}, right panel. Although in this case its polarization structure selects a preferred spatial direction, as expected for the linear sector, the associated particle number density and gravitational potential remain spherically symmetric, so the configuration is spherically symmetric according to the definition adopted in this work.

\subsubsection{Energy considerations}

In the right panel of Fig.~\ref{fig:NvsE}, we show the Newtonian energy $\mathcal{E}$ as a function of the particle number for the radially and linearly polarized Proca stars in the fundamental branch. An important feature of this diagram is that, at fixed particle number, the linearly polarized configurations always possess lower energy than the radially polarized ones.
As in the relativistic case, this  suggests that the linear branch constitutes the ground-state family of the Newtonian theory, even though both the nodeless linear and radially polarized configurations are stable. In contrast to the relativistic case, however, this conclusion can be established rigorously~\cite{Nambo:2024hao}. In the Newtonian theory, it has been shown that the energy functional~(\ref{eq.energy}) is bounded from below at fixed particle number, and that its minimum is attained by the fundamental branch of the linear polarization states. Moreover, owing to the enhanced $U(3)$ symmetry of the Newtonian system, this ground state is degenerate, and any configuration related to a linearly polarized solution by a global unitary transformation possesses exactly the same particle number, density profile, gravitational potential, and total energy, and therefore also qualifies as a ground state. This includes, for instance, circularly and, more generally, elliptically polarized configurations.
Throughout this work, however, we will not distinguish between solutions related in this way and will regard them as belonging to the same ground-state family.

%%%%%%%%%%%%%%%%%%%%%%%%%%%%%%%%%%%%%%%%%%%%%%%%%%%%%%%%%%%%%%%%%%%%%%%%%%%%%%
\section{Relativistic and Newtonian Proca stars: a comparison}\label{sec.relativistic_newtonian_conection}
%%%%%%%%%%%%%%%%%%%%%%%%%%%%%%%%%%%%%%%%%%%%%%%%%%%%%%%%%%%%%%%%%%%%%%%%%%%%%%

In the previous section, we constructed Proca stars as classical solutions of the Einstein-Proca theory and of its Newtonian limit, described by the \mbox{spin-1} Schr\"odinger-Poisson system. We now turn to a detailed comparison between the two descriptions, focusing on the relativistic solutions in the limit $\omega \to \mu$, where the Newtonian approximation is expected to be valid. At first sight, this comparison reveals apparent inconsistencies between the two approaches (see Table~\ref{table}). However, a more careful analysis shows that, once the relevant scalings and identifications are properly taken into account, the relativistic and Newtonian descriptions are in full agreement in this regime.

To carry out this comparison, we must first establish the relation between the dimensionless variables employed in the relativistic and Newtonian numerical codes. This can be achieved by comparing Eqs.~\eqref{eq.dimensionless.relativistic} and~\eqref{eq.dimensionless.Newtonian}, and using the fact that, in the Newtonian limit, the spacetime metric can be written as
\begin{subequations}
\begin{equation}
g_{00\textrm{phys}} = -(1+2\mathcal{U}_{\textrm{phys}}),\quad g_{ij\textrm{phys}}=(1-2\mathcal{U}_{\textrm{phys}})\delta_{ij}
\end{equation}
while the components of the Proca field satisfy\footnote{We denote by $\vec{\mathcal{A}}$ the spatial components of the vector field obtained by raising the index of the one-form $\mathcal{A}$.}
\begin{equation}
\mathcal{A}_{0\textrm{phys}}=\frac{1}{\sqrt{\mu^3}}e^{-\left(i\mu t_{\textrm{phys}}-\frac{\pi}{2}\right)}\nabla_{\textrm{phys}}\cdot\vec{\psi}_{\textrm{phys}},\quad 
\vec{\mathcal{A}}_{\textrm{phys}} = \frac{1}{\sqrt{\mu}}e^{-i\mu t_{\textrm{phys}}}\vec{\psi}_{\textrm{phys}},
\end{equation}
\end{subequations}
with $\omega_{\textrm{phys}} = \mu + E_{\textrm{phys}}$. Combining these relations, one obtains the following correspondence between relativistic and Newtonian code variables:
\begin{subequations}\label{eq.comparisons}
\begin{eqnarray}
& \displaystyle{g_{00\textrm{R}}+1 = -\left(\lambda_*\frac{m_{\textrm{Pl}}^2}{\mu^2}\right)^{-1}\mathcal{U}_{\textrm{N}},\quad g_{ij\textrm{R}}-\delta_{ij} = -\left(\lambda_*\frac{m_{\textrm{Pl}}^2}{\mu^2}\right)^{-1}\mathcal{U}_{\textrm{N}}\delta_{ij},}\\
& \displaystyle{\mathcal{A}_{0\textrm{R}}=\left( \lambda_* \frac{m_{\textrm{Pl}}^2}{\mu^2}\right)^{-3/2}e^{-\left(i t_{\textrm{R}}-\frac{\pi}{2}\right)}\nabla\cdot\frac{\vec{\psi}_{\textrm{N}}}{\sqrt{2}},  \quad 
\vec{\mathcal{A}}_{\textrm{R}} = \left(\lambda_*\frac{m_{\textrm{Pl}}^2}{\mu^2}\right)^{-1}e^{-it_{\textrm{R}}}\frac{\vec{\psi}_{\textrm{N}}}{\sqrt{2}}},\\
& \displaystyle{\omega_{\textrm{R}} - 1 = \left(\lambda_*\frac{m_{\textrm{Pl}}^2}{\mu^2}\right)^{-1}\frac{E_{\textrm{N}}}{2}},\label{eq.comparison.omega}
\end{eqnarray}
where again $m_{\textrm{Pl}}^{-2}=8\pi G$. It should be understood that these relations apply to spatially dependent quantities evaluated at corresponding points, with the coordinates related through the scaling
\begin{equation}\label{eq.comparisons.legth}
\vec{x}_{\textrm{R}} = \left( \lambda_* \frac{m_{\textrm{Pl}}^2}{\mu^2}\right)^{1/2}\vec{x}_{\textrm{N}}.
\end{equation}
\end{subequations}
This correspondence is expected to hold only within the regime of validity of the Newtonian approximation. Note that relativistic and Newtonian quantities are related by appropriate powers of the single dimensionless combination $\lambda_* m_{\textrm{Pl}}^2/\mu^2$, and since $\lambda_*$ is arbitrary, these relations allow the conversion to be carried out for any value of the Proca mass $\mu$. 

In practice, comparing a relativistic solution in the limit $\omega_{\textrm{R}}\to 1$ with its Newtonian counterpart reduces to fixing the scaling factor $\lambda_* m_{\textrm{Pl}}^2/\mu^2$ by matching a single pair of corresponding physical observables. For practical purposes, it is most convenient to use a global quantity that characterizes the solutions. For definiteness, we shall employ the frequency relation given in Eq.~\eqref{eq.comparison.omega}. Once this matching is performed, all remaining quantities, such as the metric or the field components, are uniquely determined through Eqs.~\eqref{eq.comparisons}.

%%%%%%%%%%%%%%%%%%%%%%%%%%%%%%%%%%%%%%%%%%%%%%%%%%%%%%%%%%%%%%%%%%%%%%%%%%%%%%%%%%%%%%%%%%
%\subsection{Correspondence between $\ell=0$ electric and radially polarized Proca stars}\label{sec:correspondece}
\subsection{Correspondence between \texorpdfstring{$\ell=0$}{l=0} electric and radially polarized Proca stars}\label{sec:correspondece}
%%%%%%%%%%%%%%%%%%%%%%%%%%%%%%%%%%%%%%%%%%%%%%%%%%%%%%%%%%%%%%%%%%%%%%%%%%%%%%%%%%%%%%%%%%

\begin{figure*}[t!]
\includegraphics[width=\textwidth]{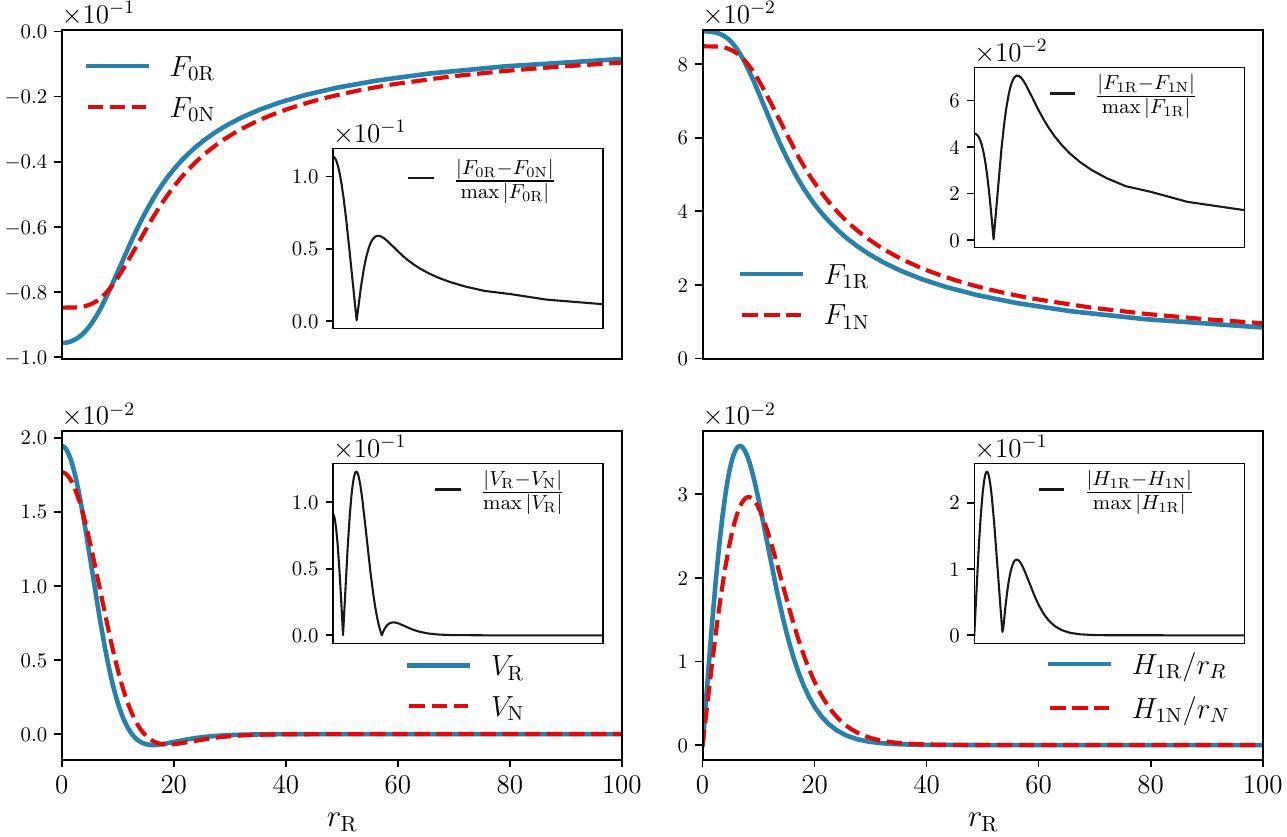}
\caption{\small\textbf{Relativistic-Newtonian comparison for an electric Proca star with $\ell=0$, $\omega_{\textrm{R}}=0.950$, and $n=0$.} Blue solid curves represent the relativistic monopolar ($\ell=0$) solution, while red dashed curves its Newtonian radially polarized counterpart. The first row displays the gravitational potentials, whereas the second one the nonvanishing vector field components. The inset table lists the relative discrepancies, normalized to the corresponding maximum values. The visible mismatch confirms that, for $\omega_{\textrm{R}}=0.950$, relativistic corrections remain significant.}\label{Fig:Spherical0.95}
\end{figure*}

\begin{figure*}[t!]
\includegraphics[width=\textwidth]{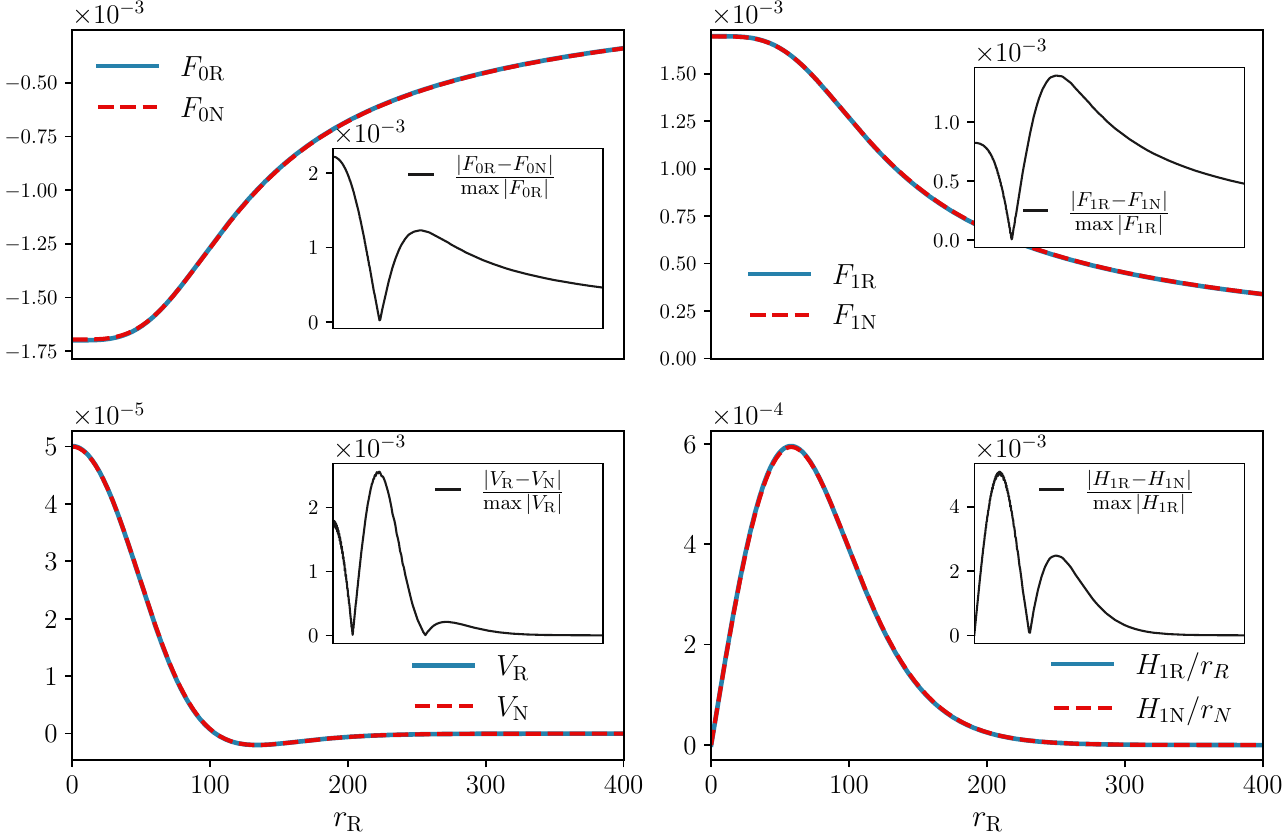}
\caption{\small\textbf{Relativistic-Newtonian comparison for an electric Proca star with $\ell=0$, $\omega_{\textrm{R}}=0.999$, and $n=0$.} Same as in  Fig.~\ref{Fig:Spherical0.95}, but for a configuration well within the Newtonian regime. In this case, all physical quantities agree remarkably well, with maximum relative errors below $0.6\, \%$.}\label{Fig:Spherical0.999}
\end{figure*}

The qualitative morphological similarities between the relativistic $\ell=0$ electric Proca stars and the Newtonian radially polarized Proca stars, visible in the left panels of Figs.~\ref{fig.relativisticPS} and~\ref{fig.NewtonianPS}, naturally motivate a quantitative comparison between these two families and suggest that the former approach the latter in the Newtonian limit. 

To test this correspondence, in Figs.~\ref{Fig:Spherical0.95} and~\ref{Fig:Spherical0.999} we compare relativistic and Newtonian solutions along the fundamental branch for configurations with $\omega_{\textrm{R}}=0.950$ and $\omega_{\textrm{R}} = 0.999$, respectively. In order to carry out such a comparison, we make use of the relations~(\ref{eq.comparisons}). For the specific correspondence between $\ell=0$ electric Proca stars and radially polarized Newtonian configurations, they reduce to
\begin{subequations}
\begin{eqnarray}
& \displaystyle{F_{0\textrm{R}} = \left(\lambda_*\frac{m_{\textrm{Pl}}^2}{\mu^2}\right)^{-1}\frac{\mathcal{U}_{\textrm{N}}}{2},\quad F_{1\textrm{R}} = - \left(\lambda_*\frac{m_{\textrm{Pl}}^2}{\mu^2}\right)^{-1}\frac{\mathcal{U}_{\textrm{N}}}{2}}, \\
&\displaystyle{ V_{\textrm{R}}=\left(\lambda_*\frac{m_{\textrm{Pl}}^2}{\mu^2}\right)^{-3/2}\frac{1}{r_{\textrm{N}}^2}\frac{\partial}{\partial r_{\textrm{N}}}\left(r_{\textrm{N}}^2\frac{\sigma_{\textrm{N}}}{\sqrt{2}}\right), \quad \frac{H_{1\textrm{R}}}{r_{\textrm{R}}} = \left(\lambda_*\frac{m_{\textrm{Pl}}^2}{\mu^2}\right)^{-1}\frac{\sigma_{\textrm{N}}}{\sqrt{2}}}.
\end{eqnarray}
\end{subequations}
It is important to note that relativistic quantities are functions of the radial coordinate $r_{\textrm{R}}$, whereas the Newtonian ones are naturally expressed in terms of $r_{\textrm{N}}$. Therefore, in order to perform a direct comparison, all quantities must be expressed in terms of a common radial variable. In practice, we rewrite the Newtonian solutions using the rescaled $r_{\textrm{R}}$, as dictated by Eq.~(\ref{eq.comparisons.legth}), before carrying out the comparison.
As shown in the first of these figures, Fig.~\ref{Fig:Spherical0.95}, the agreement between the relativistic and Newtonian profiles is still poor at $\omega_{\textrm{R}}=0.950$, indicating that the solution is still far from the Newtonian regime and that relativistic corrections remain significant. By contrast, Fig.~\ref{Fig:Spherical0.999} shows excellent agreement across all physical quantities, with the differences becoming visually indistinguishable at the scale of the plots. This provides strong evidence that radially polarized Proca stars arise as the Newtonian limit of the relativistic $\ell=0$ electric solutions.

\begin{figure}[t!]
\centering
\includegraphics[width=\textwidth]{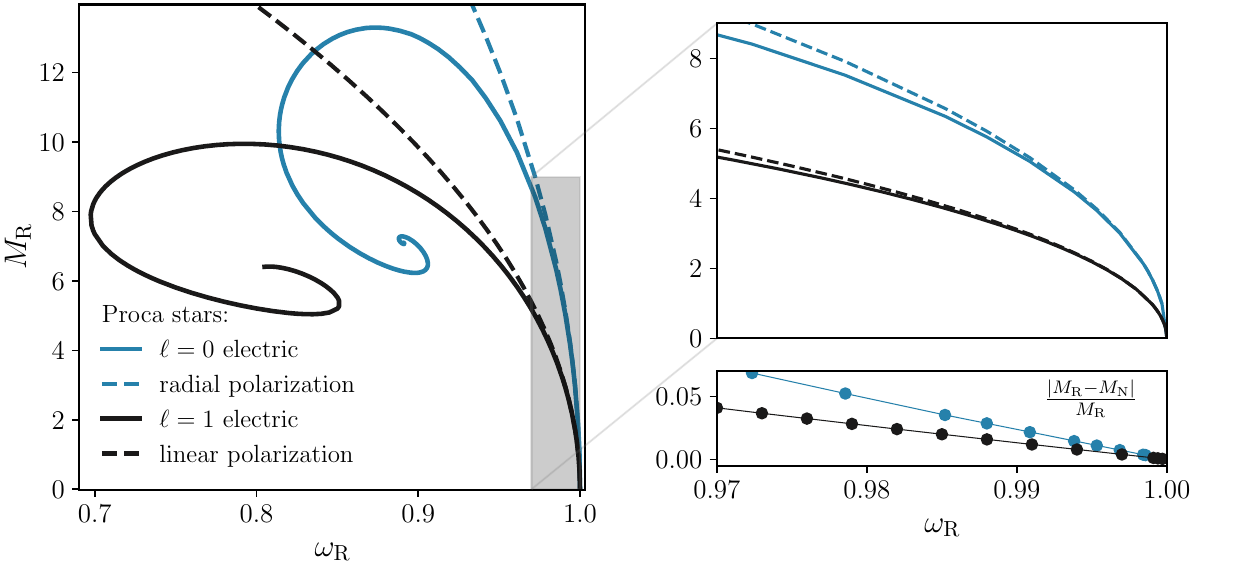}
\caption{\small\textbf{Mass-frequency diagram: relativistic-Newtonian correspondence.} Same mass-frequency diagrams as in Fig.~\ref{fig:pspace}, after rescaling the Newtonian solutions according to Eqs.~(\ref{eq.comparisons}). The secondary panels on the right show that the two descriptions converge in the limit $\omega_{\textrm{R}} \to 1$, as expected from the Newtonian approximation. In particular, for $\omega_{\textrm{R}} \geq 0.980$, the relative error remains below $5\%$.}\label{fig:equivalenceMass}
\end{figure}

\begin{figure}[t!]
\centering
\includegraphics[width=\textwidth]{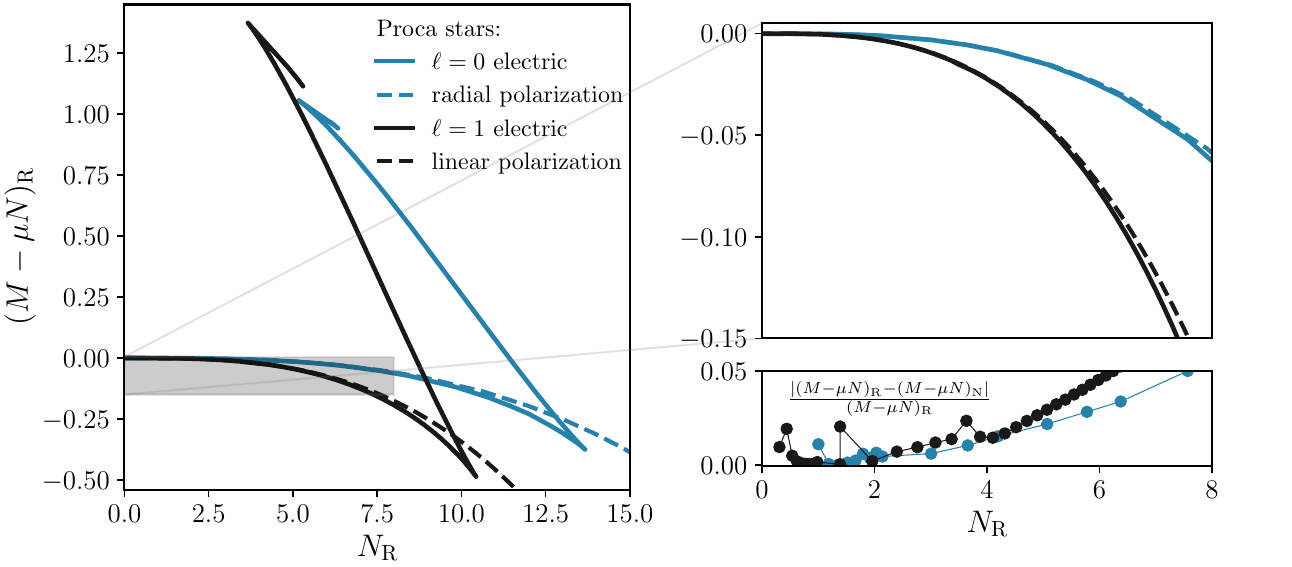}
\caption{\small\textbf{Energy-particle number diagram: relativistic-Newtonian correspondence.} Same energy-particle number diagrams as in Fig.~\ref{fig:NvsE}, after rescaling the Newtonian solutions according to Eqs.~(\ref{eq.comparisons}). The secondary panels on the right show that the two descriptions converge in the limit $\omega_{\textrm{R}} \to 1$ (i.e., $N_R\to 0$), as expected from the Newtonian approximation. In particular, for $\omega_{\textrm{R}} \geq 0.975$, the relative error remains below $5\%$.}\label{fig:equivalenceEnergy}
\end{figure}

This interpretation is further supported by Fig.~\ref{fig:equivalenceMass}, where we compare the corresponding rescaled families of solutions through the mass-frequency diagram shown in Fig.~\ref{fig:pspace}. As $\omega_{\textrm{R}}\to 1$, the relative difference between the two descriptions tends to zero, while already for $\omega_{\textrm{R}}\gtrsim 0.970$ the discrepancy remains below $7\%$. The corresponding comparison based on the energy-particle number diagram of Fig.~\ref{fig:NvsE} is presented in Fig.~\ref{fig:equivalenceEnergy}, where a similar level of agreement is observed. Combined with the close match between the field profiles, these results provide compelling evidence that the relativistic monopolar branch approaches the Newtonian hedgehog family in the nonrelativistic limit.

\begin{figure}[t!]
\centering
\includegraphics[width=0.5\textwidth]{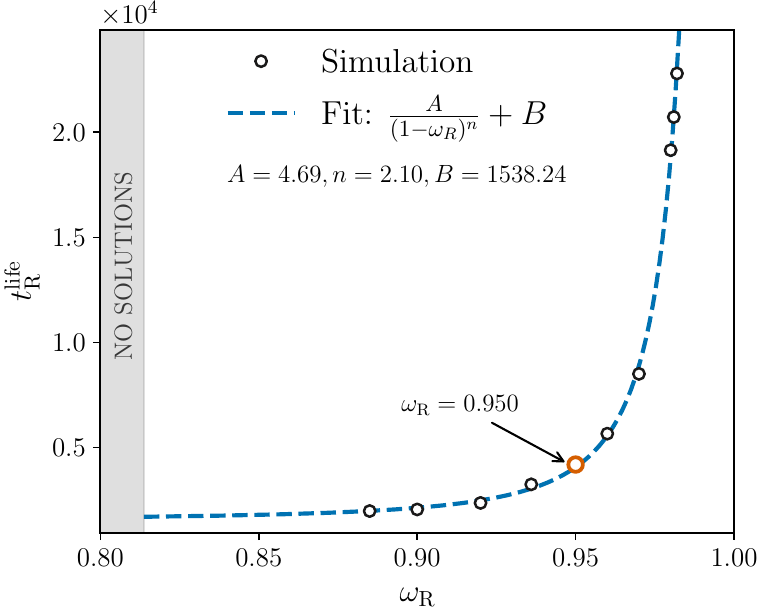}\hspace{0.5cm}\includegraphics[width=0.45\textwidth]{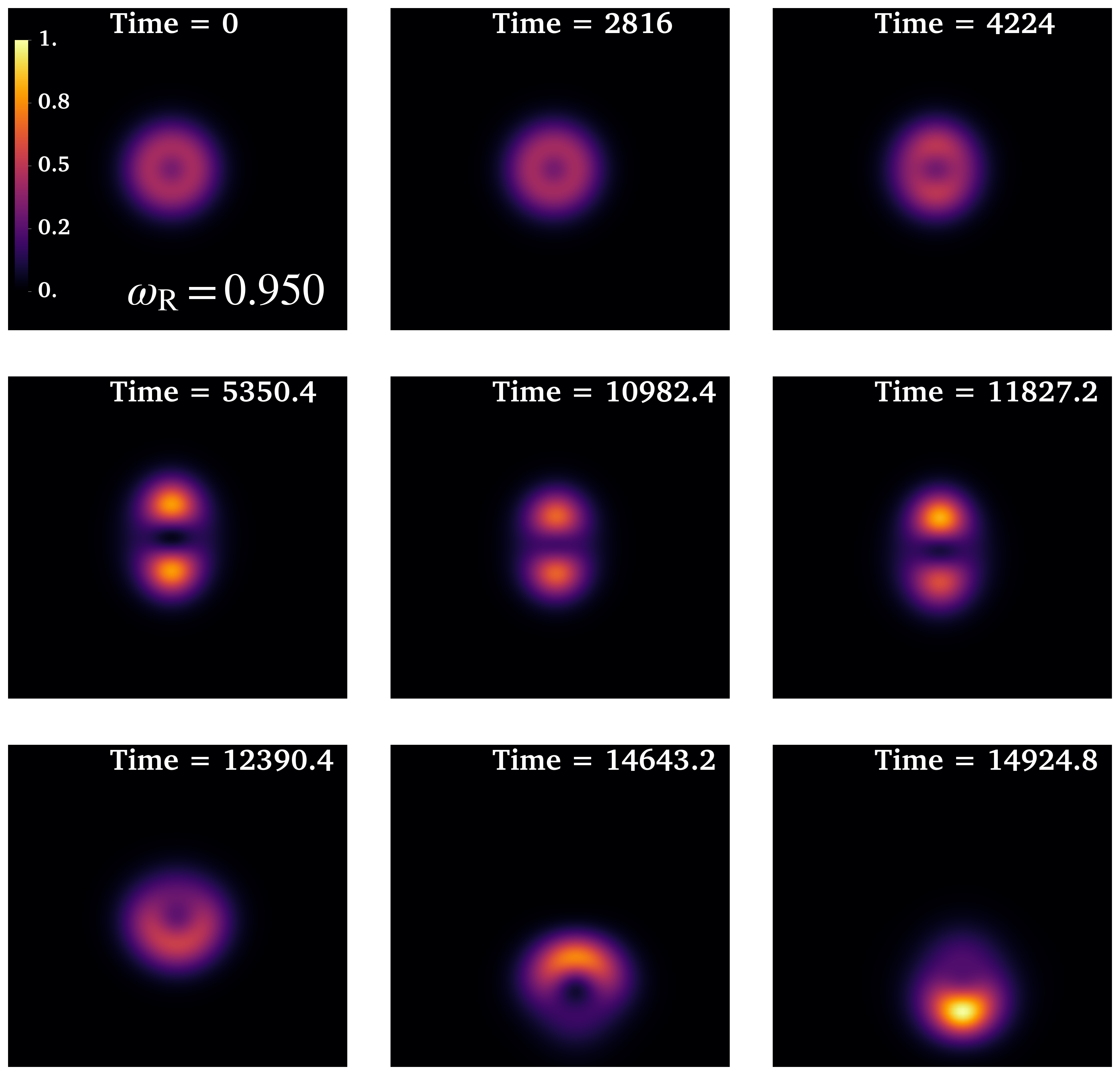}
\caption{\small\textbf{Lifetime of $\ell=0$ electric Proca stars.} \textit{Left panel:} Lifetime as a function of frequency. The markers correspond to fully nonlinear dynamical evolutions, while the dashed curve shows the best-fit model used to extrapolate the behavior toward the Newtonian limit. \textit{Right panel:} Snapshots of the time evolution of the Proca Komar energy density in the $xz$ plane for the spherically symmetric Proca star with $\omega_{\mathrm{R}}=0.950$ (indicated by the red marker in the left panel). The full movie is available in Ref.~\cite{youtube}. The energy density has been normalized to its maximum value during the evolution. Time increases from left to right and from top to bottom. The instability manifests itself through the loss of the initial shell-like spherical structure and the subsequent development of a prolate configuration. The onset of this transition occurs between $t_{\mathrm{R}}=4000$ and $4300$. 
}\label{fig:stability}
\end{figure}

Despite the clear agreement between the two families in the Newtonian regime, an apparent discrepancy remains at the level of stability: $\ell=0$ electric Proca stars are unstable under generic perturbations~\cite{Herdeiro:2023wqf}, while the corresponding Newtonian radially polarized states are linearly stable in the absence of radial nodes~\cite{Nambo:2025lnu}. To investigate this issue, we have performed fully nonlinear time evolutions of $\ell=0$ configurations along the fundamental branch as the frequency $\omega_{\textrm{R}}$ approaches unity. The development of the instability is monitored through the evolution of the energy density distribution, with the onset of the unstable phase signaled by the loss of the initial shell-like spherical structure of the configuration. This deformation is attributed to the growth of a non-spherical unstable mode, which, upon reaching sufficient amplitude for nonlinear effects to become important, drives the subsequent transition toward a prolate configuration, as illustrated in the right panel of Fig.~\ref{fig:stability} for the case $\omega_{\mathrm{R}}=0.950$. 
No explicit perturbation is added to the initial data, so that the instability is seeded by numerical truncation errors associated with the discretized evolution.
See App.~\ref{app.evol} for additional details on the numerical setup, convergence test, and a dynamical analysis of the Proca star polarization. In particular, Fig.~\ref{fig:3d_evovolution} shows snapshots of the evolution of the configuration with $\omega_{\mathrm{R}}=0.900$, including the polarization vector, where the migration toward a configuration with quasi-linear polarization can be seen more clearly.

As illustrated in the left panel of Fig.~\ref{fig:stability}, the lifetime of these configurations, measured from the start of the simulation until the onset of the instability,
increases as the Newtonian limit is approached, and our results are consistent with it 
becoming arbitrarily large in the limit $\omega_{\textrm{R}} \to 1$. To quantify this behavior, we fitted the numerical data to the function
\begin{equation}
t_{\textrm{R}}^{\textrm{life}} = \frac{A}{(1-\omega_{\textrm{R}})^n} + B, 
\end{equation}
obtaining $A=4.69$, $n=2.10$, and $B=1538.24$, with a coefficient of determination $R^2=0.9993$. The excellent agreement between the fit and the data over the range considered is consistent with a rapidly increasing instability onset time toward the Newtonian limit. Since the instability is seeded by numerical truncation errors, however, the absolute value of the onset time may depend on the numerical resolution and should not be interpreted as a direct measurement of the linear growth rate. Rather, these measurements are intended to characterize the timescale over which the instability becomes dynamically relevant and, in particular, its systematic increase as the Newtonian limit is approached.

To further validate this interpretation, in App.~\ref{app.evol} we analyze the unstable mode in the evolution of the $\ell=0$ Proca star with $\omega_{\rm R}=0.900$.
Although truncation errors excite the first three azimuthal modes of the energy density with comparable amplitudes, the $m=2$ mode is the dominant unstable mode, as shown in the left panel of Fig.~\ref{fig:unstable_mode}. Its exponential growth provides an independent estimate of the instability timescale and, consequently, of the corresponding instability lifetime. For this configuration, the two timescales, the first inferred from the loss of the initially spherical configuration and the second extracted independently from the growth of the dominant unstable mode, are in excellent agreement. Moreover, the growth rate extracted at three different grid resolutions is essentially unchanged, as shown in the right panel of Fig.~\ref{fig:unstable_mode}. This demonstrates that the extracted growth rate is insensitive to the numerical discretization at the resolutions considered, supporting the interpretation that the observed growth is dominated by the physical unstable mode rather than by numerical errors. Further details are given in App.~\ref{app.evol}.

At this point, the mechanism responsible for the stabilization of nodeless $\ell=0$ configurations in the Newtonian theory remains poorly understood. We conjecture that it is related to the symmetry enhancement characteristic of the nonrelativistic regime. In this picture, the instability of the relativistic $\ell=0$ branch would become progressively weaker as the Newtonian limit is approached and ultimately disappear in the strict Newtonian limit, consistently with the linear stability of the corresponding radially polarized solutions of the spin-1 Schr\"odinger-Poisson system. One possible explanation is that the variational structure of the problem changes in the presence of the enlarged $U(3)$ symmetry. In particular, $\ell=0$ configurations may correspond to saddle points (or even local maxima) of the energy functional when only the particle number $N$ is held fixed, while becoming local minima once the full set of conserved quantities associated with the $U(3)$ symmetry, encoded in the Hermitian tensor $\hat{Q}$ of Eq.~(\ref{eq.hat.Q}), is taken into account. If correct, this would provide a natural explanation for the disappearance of the instability in the Newtonian regime. A preliminary investigation of this problem, in the context of $\ell$-boson stars and restricted to spherical perturbations, was recently carried out in~\cite{Nambo:2026yfd}. At present, however, a rigorous understanding of the stabilization mechanism is still lacking.

%%%%%%%%%%%%%%%%%%
%\subsection{Correspondence between $\ell=1$ electric and linearly polarized Proca stars}
\subsection{Correspondence between \texorpdfstring{$\ell=1$}{l=1} electric and linearly polarized Proca stars}
%%%%%%%%%%%%%%%%%%

Although there are no {\it a priori} morphological arguments suggesting a direct correspondence between the relativistic $\ell=1$ electric Proca stars and the Newtonian linearly polarized configurations, the identification of the latter as the ground-state family of the Newtonian theory, together with evidence that the former may play an analogous role in the relativistic theory, naturally motivates such a comparison.

\begin{figure*}[t!]
\includegraphics[width=\textwidth]{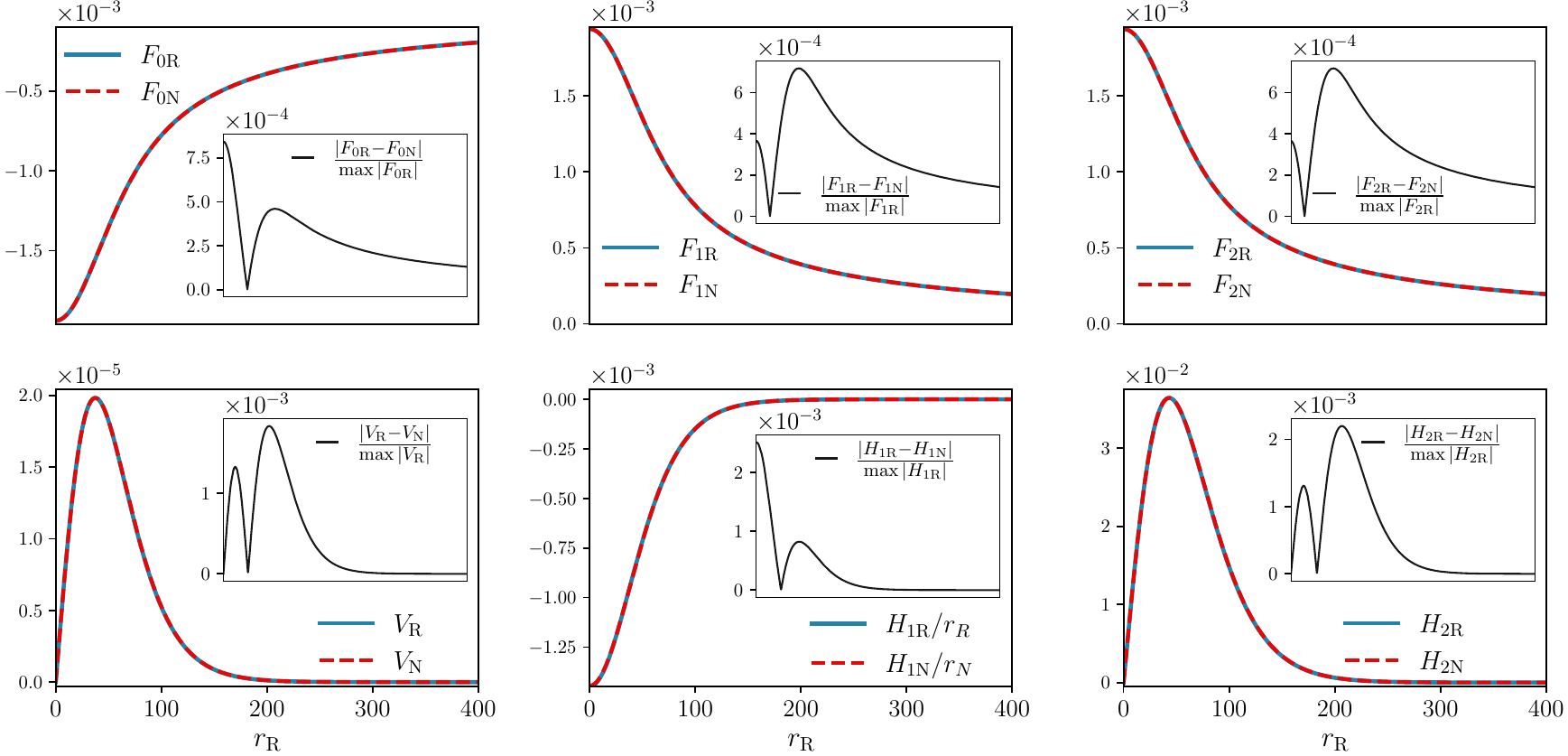}
\caption{\small\textbf{Relativistic-Newtonian comparison for an electric Proca star with $\ell=1$, $\omega_{\textrm{R}}=0.999$, and $n=0$.} Same as in  Fig.~\ref{Fig:Spherical0.999}, but for a configuration with $\ell =1$. All physical quantities agree remarkably well, with maximum relative errors below $0.3\, \%$.}\label{Fig:Prolate0.999n=0}
\end{figure*}

\begin{figure*}[t!]
\includegraphics[width=\textwidth]{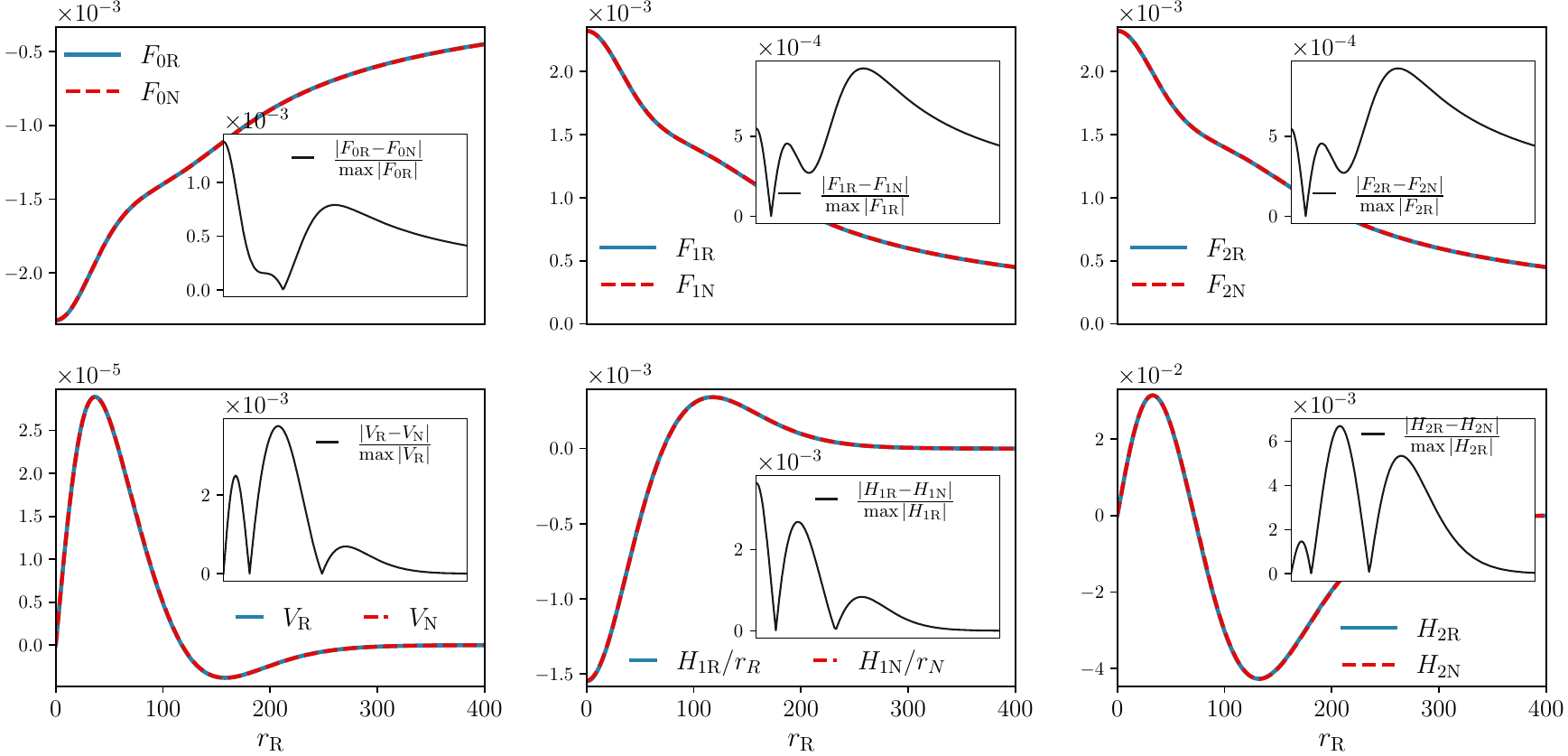}
\caption{\small\textbf{Relativistic-Newtonian comparison for an electric Proca star with $\ell=1$, $\omega_{\textrm{R}}=0.999$, and $n=1$.} Same as in  Fig.~\ref{Fig:Prolate0.999n=0}, but for a configuration with one node. All physical quantities agree remarkably well, with maximum relative errors below $0.7\, \%$.}\label{Fig:Prolate0.999n=1}
\end{figure*}

To test this correspondence, in Fig.~\ref{Fig:Prolate0.999n=0} we compare relativistic and Newtonian solutions along the fundamental branch for a configuration with $\omega_{\textrm{R}}=0.999$. 
Again, to carry out such a comparison, we make use of the relations~(\ref{eq.comparisons}). For the specific correspondence between $\ell=1$ electric Proca stars and linearly polarized Newtonian configurations, they reduce to
\begin{subequations}
\begin{eqnarray}
& \displaystyle{F_{0\textrm{R}} = \left(\lambda_*\frac{m_{\textrm{Pl}}^2}{\mu^2}\right)^{-1}\frac{\mathcal{U}_{\textrm{N}}}{2},\quad F_{1\textrm{R}} = - \left(\lambda_*\frac{m_{\textrm{Pl}}^2}{\mu^2}\right)^{-1}\frac{\mathcal{U}_{\textrm{N}}}{2},\quad F_{2\textrm{R}} = - \left(\lambda_*\frac{m_{\textrm{Pl}}^2}{\mu^2}\right)^{-1}\frac{\mathcal{U}_{\textrm{N}}}{2}}, \\
&\displaystyle{ \frac{V_{\textrm{R}}}{\cos\theta}=\left(\lambda_*\frac{m_{\textrm{Pl}}^2}{\mu^2}\right)^{-3/2}\hspace{-.25cm}\frac{\partial}{\partial r_{\textrm{N}}}\left(\frac{\sigma_{\textrm{N}}}{\sqrt{2}}\right), \;\;\frac{H_{1\textrm{R}}}{r_{\textrm{R}}\cos\theta} = \left(\lambda_*\frac{m_{\textrm{Pl}}^2}{\mu^2}\right)^{-1}\hspace{-.25cm}
\frac{\sigma_{\textrm{N}}}{\sqrt{2}},\;\; \frac{H_{2\textrm{R}}}{\sin\theta} = -\left(\lambda_*\frac{m_{\textrm{Pl}}^2}{\mu^2}\right)^{-1/2}\hspace{-.3cm}r_{\textrm{N}}}\frac{\sigma_{\textrm{N}}}{\sqrt{2}}.\quad\;\;
\end{eqnarray}
\end{subequations}
As in the $\ell=0$ electric case (see Fig.~\ref{Fig:Spherical0.999}), the agreement is excellent across all physical quantities, with the differences becoming visually indistinguishable at the scale of the plots. 
For completeness, in Fig.~\ref{Fig:Prolate0.999n=1} we repeat the same comparison for the first excited family ($n=1$) at $\omega_{\textrm{R}}=0.999$, again finding excellent agreement. 
This correspondence is further supported by Figs.~\ref{fig:equivalenceMass} and~\ref{fig:equivalenceEnergy}, where the appropriately rescaled families of solutions extracted from Figs.~\ref{fig:pspace} and~\ref{fig:NvsE} are directly compared. 
As these figures demonstrate, the relativistic and Newtonian descriptions become increasingly consistent as the limits $\omega_{\mathrm{R}}\to1$ and $N_{\mathrm{R}}\to0$ are approached. In particular, the corresponding mass-frequency and energy-particle number relations asymptotically coincide in the Newtonian regime.

\begin{figure*}[t!]
\includegraphics[width=\textwidth]{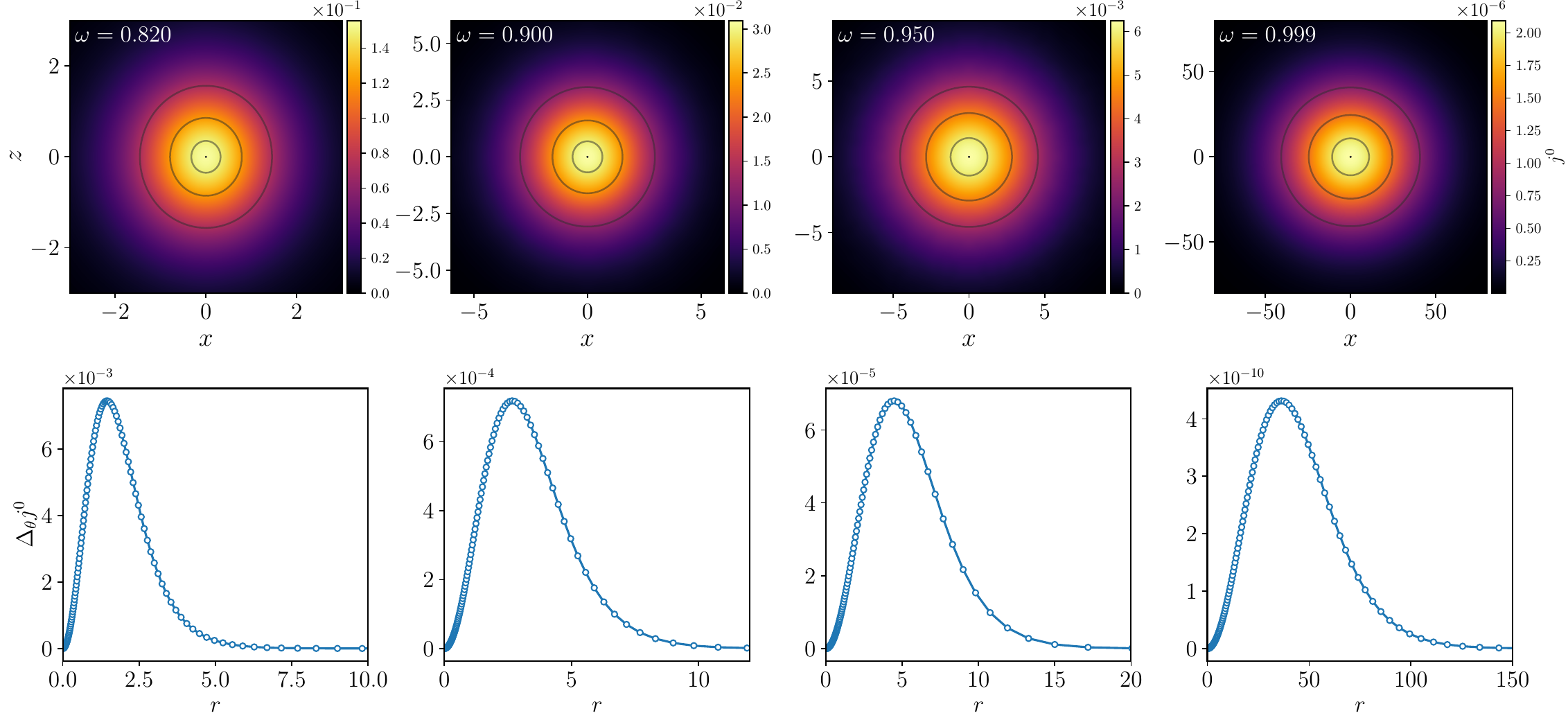}
\caption{\small\textbf{Angular structure of the particle number density $j^0$ for $\ell=1$ electric Proca stars.} {\it Top panels:} Spatial distribution of $j^0$ in the $xz$ meridional plane for different values of $\omega$. Color shading represents the particle number density, while solid lines denote isodensity contours. {\it Bottom panels:} Radial profile of the angular spread, $\Delta_\theta j^0(r) = \max_\theta j^0 - \min_\theta j^0$.  As the Newtonian regime is approached, the configurations become progressively more spherical.}\label{fig.eccentricity}
\end{figure*}

This provides strong evidence that linearly polarized Proca stars arise as the Newtonian limit of the relativistic $\ell=1$ electric solutions. Nevertheless, an apparent discrepancy remains in the morphology of the configurations: while relativistic $\ell=1$ electric Proca stars exhibit a characteristic prolate deformation~\cite{Herdeiro:2023wqf}, the corresponding Newtonian linearly polarized states are spherically symmetric~\cite{Nambo:2024hao}. This suggests that the prolateness of the relativistic configurations is a genuine relativistic effect, which gradually disappears as the Newtonian limit is approached. To verify this behavior, in Fig.~\ref{fig.eccentricity} we analyze the angular dependence of the particle number density for different frequencies. As the figure shows, the configurations become progressively more spherical as the frequency approaches unity, smoothly approaching the morphology of their Newtonian counterparts.

The correspondence between relativistic $\ell=1$ electric Proca stars and Newtonian linearly polarized states further supports the interpretation that the former constitute the ground-state family of the Einstein-Proca theory, at least in the Newtonian regime.  There is, however, an important caveat that should be noted.  In the Newtonian theory, the ground state is degenerate: linearly polarized configurations belong to a broader family of constant polarization states related by unitary 3×3 transformations, all of which possess the same particle number, density profile, gravitational potential, and total energy. This degeneracy is a direct consequence of the $U(3)$ invariance of the spin-1 Schr\"odinger-Poisson system and is not expected to persist once relativistic corrections are taken into account. Consequently, configurations that are equivalent in the Newtonian theory need not remain energetically degenerate in the relativistic regime. It therefore remains possible that, if the Einstein-Proca theory admits a ground-state family, a different representative of the degenerate Newtonian ground state, rather than the $\ell=1$ electric branch, becomes energetically preferred away from the Newtonian limit.

While this possibility cannot presently be excluded, there is currently no evidence that any other relativistic branch becomes energetically preferred over the $\ell=1$ electric solutions. On the contrary, the dynamical stability of the $\ell=1$ electric Proca stars provides further support for the possibility that they constitute the ground-state family of the Einstein-Proca system, corresponding to the lowest-energy configurations at fixed particle number among all configurations admitted by the theory. A definitive answer, however, would require a rigorous variational proof establishing that these solutions indeed minimize the energy under variations that preserve the particle number, which is presently unavailable.

%%%%%%%%%%%%%%%%%%%%%%%%%%%%%%%%%%%%%%%%%%%%%%%%%%%%%%%%%%%%%%%%%%%%%%%%%%%%%%%%%%%%%
\section{Conclusions}\label{sec.conclusions}
%%%%%%%%%%%%%%%%%%%%%%%%%%%%%%%%%%%%%%%%%%%%%%%%%%%%%%%%%%%%%%%%%%%%%%%%%%%%%%%%%%%%%

In this paper, we have presented a comprehensive comparison between relativistic Proca stars, obtained as self-gravitating solutions of the Einstein-Proca equations, and Newtonian Proca stars, described by the spin-1 Schr\"odinger-Poisson system. Particular attention has been devoted to the regime $\omega\to\mu$, where the Newtonian approximation provides the nonrelativistic  description of the Einstein-Proca theory. Our analysis has focused on establishing the correspondence between these two descriptions and clarifying the similarities and differences between their respective solution spaces.

For Proca fields, the correspondence between the relativistic and Newtonian limits is not straightforward. The vector nature of the field gives rise to distinct polarization structures and, in the nonrelativistic regime, to a symmetry enhancement of the theory, leading to a considerably richer solution space. As a result, the mapping between relativistic and Newtonian solutions must be determined explicitly. Our analysis establishes the following correspondence:
\begin{itemize}
    \item radially polarized Proca stars arise as the Newtonian limit of $\ell=0$ electric Proca stars,
    \item linearly polarized Proca stars arise as the Newtonian limit of $\ell=1$ electric Proca stars.
\end{itemize}

This identification carries nontrivial physical implications. In particular, the fundamental branch of linearly polarized Proca stars is known to realize the ground-state family of the Newtonian theory. This naturally suggests that the stable $\ell=1$ electric branch may play an analogous role in the relativistic theory, at least in the Newtonian limit of the Einstein-Proca system, in agreement with the interpretation previously put forward in Ref.~\cite{Herdeiro:2023wqf}.  
A note of caution is nevertheless warranted. At present, no rigorous mathematical result establishes the existence of a ground state in the Einstein-Proca theory. Moreover, the Newtonian ground state is highly degenerate, and linearly polarized configurations belong to a broader family of states related by global $U(3)$ transformations, including circularly and, more generally, elliptically polarized configurations. All members of this family possess the same particle number, density profile, gravitational potential, and total energy, and therefore qualify equally as ground state configurations. Since this degeneracy originates from a symmetry that is absent in the full Einstein-Proca theory, it may be lifted by relativistic corrections away from the Newtonian limit, and in that case, a different member of the degenerate Newtonian family could emerge as the true ground state of the relativistic theory. Although the existence of a dynamically stable $\ell=1$ electric branch provides evidence that the linearly polarized family constitutes the relevant continuation of the ground state into the relativistic regime, establishing whether this branch, or another member of the corresponding relativistic solution space, realizes the absolute energy minimum at fixed particle number remains an important open problem.

Our analysis also resolves several apparent discrepancies between the relativistic and Newtonian descriptions. In particular, $\ell=1$ electric Proca stars exhibit a prolate energy density distribution in the relativistic regime, whereas their Newtonian counterparts are spherically symmetric, even though the Proca field itself retains a preferred polarization direction. We have shown that the prolate deformation is progressively suppressed as the Newtonian limit is approached, demonstrating that it is an intrinsically relativistic feature. Furthermore, in the limit $\omega\to\mu$, we have also established that the relativistic solutions converge smoothly to the spherically symmetric linearly polarized configurations of the spin-1 Schr\"odinger-Poisson system.

A similar situation arises in the $\ell=0$ sector. While relativistic $\ell=0$ electric Proca stars are known to be unstable under generic perturbations, their Newtonian counterparts (the radially polarized states) are linearly stable. We have shown that the timescale associated with the relativistic instability increases as the Newtonian regime is approached, suggesting that the instability is progressively suppressed in this limit. This behavior provides a natural explanation for the stability of the corresponding Newtonian configurations and further supports the correspondence established in this work.

Finally, our results indicate that the spin-1 Schr\"odinger-Poisson system should not be viewed merely as a simplified approximation of the relativistic Einstein-Proca equations, but rather as a regime with qualitatively distinct features of its own. While the only dynamically stable Proca stars currently known in the relativistic theory belong to the $\ell=1$ electric branch, the spin-1 Schr\"odinger-Poisson system admits a much richer spectrum of stable equilibrium configurations, including radially polarized states and multi-frequency solutions. We believe that this enlarged solution space is intimately connected to the $U(3)$ symmetry that emerges in the nonrelativistic limit. Understanding how this symmetry shapes the structure, stability, and dynamics of self-gravitating vector solitons, and clarifying whether some of the associated Newtonian configurations possess undiscovered relativistic counterparts, remain particularly intriguing directions for future investigation.

%%%%%%%%%%%%%%%%%%%%%%%%%%%%%%%%%
\section*{Acknowledgments}
%%%%%%%%%%%%%%%%%%%%%%%%%%%%%%%%%

We thank Olivier Sarbach for many useful discussions and for his collaboration in the early stages of this work.
A.D.T. acknowledges support from SECIHTI-SNII and DAIP Project No. CIIC 301/2026. C.L. acknowledges support from the Generalitat Valenciana through a Santiago Grisolía Grant (CIGRIS/2022/164). A.A.R. acknowledges support from SECIHTI-SNII and from the postdoctoral fellowship ``Estancias Posdoctorales por M\'exico para la Formaci\'on y Consolidaci\'on de las y los Investigadores por M\'exico.'' E.S.C.F. is partially supported by the FCT grant PRT/BD/153349/2021 (\url{https://doi.org/10.54499/PRT/BD/153349/2021}) under
the IDPASC Doctoral Program and by CNPq/PDJ 153723/2025-4. N.S.G. acknowledges support from the Spanish Ministry of Science, Innovation, and Universities via the Ram\'on y Cajal programme (grant RYC2022-037424-I), funded by  MICIU/AEI/10.13039/501100011033 and by ESF+.
This work is supported by CIDMA (\url{https://ror.org/05pm2mw36}) under the Portuguese Foundation for Science and Technology (FCT, \url{https://ror.org/00snfqn58}) Grants UID/04106/2025 
(\url{https://doi.org/10.54499/UID/04106/2025}) and UID/PRR/04106/2025 (\url{https://doi.org/10.54499/UID/PRR/04106/2025}), 
the projects: Horizon Europe staff exchange (SE) programme HORIZON-MSCA2021-SE-01 Grant No. NewFunFiCO-101086251 and  2022.04560.PTDC (\url{https://doi.org/10.54499/2022.04560.PTDC}),
and the Spanish Agencia Estatal de Investigaci\'on (Grant PID2024-159689NB-C21) funded by  MICIU/AEI/10.\allowbreak 13039/501100011033 and by FEDER/EU. The authors acknowledge computer resources provided by the Departamento de Física de la Universidad de Guanajuato (COUGHS from the UGDataLab), the Red Espa\~nola de Supercomputaci\'on (Tirant, MareNostrum5, Altamira, and Storage5), the technical support from the IT departments of the Universitat de Val\`encia and the Barcelona Supercomputing Center (Projects No.~RES-FI-2024-2-0012 and No.~RES-FI-2024-3-0007), and by Instituto de Física de Cantabria  (IFCA) (Altamira) through Project No.~FI-2025-1-0011. Computational resources were also provided via FCT through project 2025.09498.CPCA.A3.

\appendix

%%%%%%%%%%%%%%%%%%%%%%%%%%%%%%%%%%%%%%%%%%%%%%%%
\section{Axisymmetric Einstein-Proca equations}\label{app.eqs}
%%%%%%%%%%%%%%%%%%%%%%%%%%%%%%%%%%%%%%%%%%%%%%%%

For completeness, in this appendix we present the explicit form of the Einstein-Proca equations corresponding to the electric ansatz~\eqref{eq.ansatz.electric} and the metric~\eqref{eq.metric}, written in terms of the dimensionless variables introduced in Eq.~\eqref{eq.dimensionless.relativistic}.

Rather than solving the Einstein equations component by component, it is convenient to consider the following linear combinations (multiplied by suitable factors of $r$ and $\sin\theta$ as to de-singularise potentially problematic terms):
\begin{subequations}
\begin{align}
-{E^{t}}_{t} + {E^{r}}_{r} + {E^{\theta}}_{\theta} -{E^{\varphi}}_{\varphi}&=0,\\[2pt]
{E^{t}}_{t}+{E^{r}}_{r}+{E^{\theta}}_{\theta}-{E^{\varphi}}_{\varphi}&=0,\\[2pt]
-{E^{t}}_{t} + {E^{r}}_{r} + {E^{\theta}}_{\theta} + {E^{\varphi}}_{\varphi}&=0,
\end{align}
\end{subequations}
where\footnote{The factor of $2$ multiplying the stress-energy tensor follows from the conventional form of the Einstein equations, with coupling $8\pi G$, together with the normalization of the Proca field introduced in Eq.~\eqref{eq.dimensionless.relativistic}.} 
\begin{align}
E_{\mu\nu}:=R_{\mu\nu}-\frac{1}{2}g_{\mu\nu}R-2T_{\mu\nu}=0.
\end{align}
These combinations are chosen such that the resulting equations take a comparatively compact form and possess a principal part resembling a Laplace-type operator, which is convenient both analytically and numerically.
One then obtains:
\begin{subequations}\label{eqs.system.electric}
\begin{align}
\mathcal{D}_{0}\left[F_{1, r}, F_{1, \theta}\right] - &rF_{1,r}-\cot\theta\,F_{1,\theta}-r(1 + rF_{2,r})F_{0,r} - (F_{2,\theta}+\cot\theta)F_{0,\theta} + H_1^2 + H_2^2\nonumber\\
&+r^2e^{-2(F_0-F_1)}V^2 + \frac{2 \mathcal{F}_{r\theta}^2}{r^2}e^{-2F_{1}}=0, \label{eqs.system.electric.1}\\[2pt]
\mathcal{D}_{F_0}[F_{2,r}, F_{2,\theta}]+&\cot\theta\,F_{2,\theta} + r F_{2, r} + r^2 F_{2,r}^2 +F_{2,\theta}^2 + r F_{0, r} + \cot\theta\,F_{0,\theta} -\frac{\mathcal{F}_{r\theta}^2}{r^2}e^{-2F_1} \nonumber\\
&+ e^{-2F_{0}}\left[(rV_{r}+\omega H_1)^2+(V_\theta+\omega H_2)^2\right]=0,\\[2pt]
\mathcal{D}_{F_2}[F_{0, r}, F_{0, \theta}] + & r^2F_{0,r}^2 + F_{0,\theta}^2 - e^{-2F_0}\left[(rV_r+\omega H_1)^2 +(V_\theta + \omega H_2)^2\right] - 2 r^2 e^{-2(F_0 - F_1)}V^2 \nonumber\\
&-\frac{\mathcal{F}_{r\theta}^2}{r^2} e^{-2F_1}=0.
\end{align}
\end{subequations}
Here we have introduced the ``field strength tensor'' $\mathcal{F}_{r\theta}:=H_{1,\theta}-rH_{2,r}$, together with the differential operator
\begin{align}
\mathcal{D}_{A}\left[f, g\right]:= e^{-A}\left[\partial_{r}\left(r^2 e^{A} f\right) + \frac{1}{\sin\theta}\partial_{\theta}\left(\sin\theta e^{A} g\right)\right].
\end{align}
which naturally appears from the elliptic structure of the equations. In particular, when $A=0$, cf. Eq.~(\ref{eqs.system.electric.1}), the operator reduces to the flat-space Laplacian written in spherical coordinates acting on axisymmetric functions.

Turning now to the Proca sector, it is convenient to enforce the Lorenz condition~\eqref{eq:lorenz} through the addition of a covariant constraint term to the action. The effective Proca Lagrangian then takes the form\footnote{The same equations can also be obtained without explicitly modifying the Lagrangian. In that case, one uses Eq.~\eqref{AEqLorentzCondition} and its derivatives to solve for the combinations $H_{1,r}$, $V_r$, and $V_\theta$, which are subsequently substituted into the temporal, radial, and angular Proca equations, respectively.}
\begin{align}\label{AEq.EffectivelyProcaLagrang}
\mathcal{L}_{\mathrm{new}}^{P}
= -\frac{1}{4}F_{\alpha\beta}^*F^{\alpha\beta}-\frac{1}{2}\mu^2\mathcal{A}^*_\alpha\mathcal{A}^\alpha
-\frac{1} {2 \xi} \nabla_\alpha \mathcal{A}^{ *\alpha } \nabla_\beta \mathcal{A}^\beta  ,
\end{align}
with $\xi=1$. Although the massive Proca theory does not possess a gauge symmetry, this procedure is analogous to the covariant gauge-fixing terms commonly introduced in Maxwell theory. In particular, for $\mu=0$, the choice $\xi=1$ reduces to the standard Feynman gauge.
Again, upon substituting the metric and vector ans\"atze into the corresponding Euler-Lagrange equations and performing straightforward algebraic manipulations, the Proca equations reduce to
\begin{subequations}\label{eqs.einstein.electric}
\begin{align}
\mathcal{D}_{F_{-}}\left[V_{,r}, V_{,\theta}\right]&=r^2e^{2F_1}\left(1-\omega^2 e^{-2F_0}\right)V  + 2\omega\left(rH_1 F_{0,r}+H_2F_{0,\theta}\right),\\[2pt]
\mathcal{D}_{K}\left[H_{1, r}, H_{1, \theta}\right] - 2 r H_{1,r} &=
2QH_2\partial_{\theta}\left(\ln[\sin\theta H_2]\right)
- H_2(rF_{+,r\theta} - 2QF_{+,\theta}) - 2rF_{1,\theta}H_{2,r} \nonumber\\
&+ H_1\left[2 + r\partial_r(2F_{+}-K) + r^2(2F_{1,r}F_{+,r}-F_{+,rr}+e^{2F_1}(1-\omega^2 e^{-2F_0}))\right]\nonumber\\
& -2r^3\omega F_{0,r}Ve^{-2(F_0-F_1)},\\[2pt]
\mathcal{D}_{K}\left[H_{2, r}, H_{2, \theta}\right] - 2 r H_{2,r} &= H_2\left[\csc^2\theta -F_{+,\theta\theta}+2F_{1,\theta}(\cot\theta+F_{+,\theta})+r^2e^{2F_1}\left(1-\omega^2e^{-2F_0}\right)\right] \nonumber\\
&-2\left(H_{1,\theta}Q-rF_{1,\theta}H_{1,r}\right)
+ H_1\left[2F_{1,\theta}-r(F_{+,r\theta}-2F_{1,\theta}F_{+,r})\right]\nonumber\\
&-2r^2\omega F_{0, \theta} V e^{-2(F_0-F_1)},
\end{align}
\end{subequations}
where we have defined
\begin{align}
F_{\pm}:=F_2 \pm F_0,\quad K:=F_{+}-2F_1,\quad Q:=rF_{1,r}+1.
\end{align}

In addition to the six equations above, we monitor the Lorenz condition~\eqref{eq:lorenz}, whose expression is given by
\begin{align}\label{AEqLorentzCondition}
\mathcal{D}_{F_{+}}\left[\tfrac{H_1}{r}, H_2\right]=e^{-2(F_0 - F_1)}r^2\omega V,
\end{align}
together with the Einstein constraint equations ${E^{r}}_{r}-{E^{\theta}}_{\theta}$ and ${E^{\theta}}_{r}$.

%%%%%%%%%%%%%%%%%%%%%%%%%%%%%%%%%%%%%%%%%%%%%%%%%%%%%%%%%%%%%%%%%%%%%%%%%%%%%%%%%%%%%%%%%%%%%%%%
%\section{Small-$r$ expansion and regularity of the $\ell=1$ solutions}\label{app.expansion}
\section{Small-\texorpdfstring{$r$}{r} expansion and regularity of the \texorpdfstring{$\ell=1$}{l=1} solutions}\label{app.expansion}
%%%%%%%%%%%%%%%%%%%%%%%%%%%%%%%%%%%%%%%%%%%%%%%%%%%%%%%%%%%%%%%%%%%%%%%%%%%%%%%%%%%%%%%%%%%%%%%%

In this appendix, we analyze the behavior of the $\ell=1$ electric Proca star solutions in the vicinity of the origin. To this end, we assume that all fields admit a regular power series expansion in the radial coordinate $r$, with coefficients that may depend on the polar angle $\theta$. Substituting these expansions into the Einstein equations~(\ref{eqs.system.electric}) and Proca equations~(\ref{eqs.einstein.electric}), and solving order by order in $r$, one obtains the most general local solution compatible with regularity at the origin and the equatorial reflection symmetry characteristic of $\ell=1$ configurations.

The leading-order behavior of the Proca field is found to be
\begin{subequations}
\begin{eqnarray}
V(r,\theta)&=& c_2 r \cos \theta+\dots,\\
H_1(r,\theta)&=& c_1 r \cos \theta+\dots,\\
H_2(r,\theta)&=& -c_1 r \sin \theta+\dots,
\end{eqnarray}
\end{subequations}
where $c_1$ and $c_2$ are arbitrary constants. Note that this angular dependence is determined by the $\ell=1$ spherical harmonic, $Y_{10}\sim\cos\theta$, and its derivative, $\partial_\theta Y_{10}\sim -\sin\theta$, reflecting the dipolar nature of the solution. On the other hand, the corresponding metric functions behave as
\begin{subequations}\label{small-r}
\begin{eqnarray}
F_0(r,\theta) &=& f_{00}+\frac{2\pi G}{3}\left[-c_1^2\mu^2 (1+3 \cos 2\theta)+e^{-2f_{00}}(c_2+c_1 \omega)^2  (2+3 \cos 2\theta)\right] r^2+\dots, \label{small-r1}
\\
F_1(r,\theta) &=& f_{10}-\frac{\pi G}{3}\left[c_1^2\mu^2+e^{-2f_{00}}(c_2+c_1 \omega)^2 \right] r^2+\dots,\: \: \label{small-r2}\\
F_2(r,\theta) &=& f_{10}-\frac{\pi G}{3}\left[c_1^2\mu^2+e^{-2f_{00}}(c_2+c_1 \omega)^2 \right] r^2+\dots, \label{small-r3}
\end{eqnarray}
\end{subequations}
with $f_{00}$ and $f_{10}$ additional integration constants.

Several features are worth noting. First, the spatial geometry remains locally isotropic through quadratic order, as reflected by the equality $F_1=F_2$ up to the order displayed above. However, the dipolar nature of the Proca field manifests itself instead through the angular dependence of $F_0$, whose $r^2\cos 2\theta$ contribution represents the leading anisotropic correction to the spacetime metric. 
Second, the functions $V$, $H_1$ and $H_2$ appearing in the electric ansatz vanish linearly with $r$. While this behavior is consistent with regularity at the origin, the latter is not entirely obvious because of the combination $H_1/r$ appearing in Eq.~(\ref{eq.ansatz.electric}). To verify regularity explicitly, we express the vector potential in local Cartesian coordinates $(x,y,z)$ and substitute the leading-order expansion above, obtaining\footnote{The Cartesian components are obtained from the standard coordinate transformation, ${\cal A}_i=(\partial x^\mu/\partial x^i){\cal A}_\mu$.}
\begin{eqnarray}\label{limit-to-r0}
\lim_{r \to 0 }{\cal A}_x=\lim_{r \to 0 }{\cal A}_y=0, \quad \lim_{r \to 0 }{\cal A}_z=c_1,
\end{eqnarray}
while the electric potential satisfies $\lim_{r \to 0}V=0$. Thus, although the solution carries $\ell=1$ angular structure away from the origin, the Proca field remains smooth at the origin and approaches the finite limit $\vec{\mathcal{A}} = c_1 \hat{e}_z$ in locally Cartesian coordinates. This behavior can be appreciated in the right panel of Fig.~\ref{fig.relativisticPS}, which displays a representative solution.

%%%%%%%%%%%%%%%%%%%%%%%%%%%%%%%%%%%%%%%%%%%%%%%%%%%% 
\section{$U(3)$ symmetry in the nonrelativistic limit}\label{app.U3}
%%%%%%%%%%%%%%%%%%%%%%%%%%%%%%%%%%%%%%%%%%%%%%%%%%%% 

In this appendix, we clarify in what sense the $U(3)$ symmetry emerges in the nonrelativistic limit of Proca theory.
We begin with the relativistic theory and show that, although it propagates three physical degrees of freedom, it does not possess a $U(3)$ symmetry acting on these degrees of freedom. We then take the nonrelativistic limit and show how the leading-order dynamics acquires an enhanced $U(3)$ symmetry.

Our starting point is the action for a free, massive, complex Proca field in flat spacetime,
\begin{equation}\label{eq.Proca}
\mathcal{S}_{\textrm{P}}=\frac{1}{c}\int \text{d}^4x \left(-\frac{1}{4}F_{\mu\nu}^*F^{\mu\nu}-\frac{\mu^2c^2}{2\hbar^2}\mathcal{A}_{\mu}^*\mathcal{A}^{\mu}\right),
\end{equation}
with $d^4x=cdtd^3x$. We
keep all factors of $c$ and $\hbar$ explicit since this will make the nonrelativistic limit particularly transparent. 
Writing the four-vector as $\mathcal{A}_\mu=(\mathcal{A}_0,\mathcal{A}_i)$, the Lagrangian density can be expressed in terms of its temporal and spatial components in the form
\begin{equation}\label{eq.Proca.expanded}
\mathcal{L}_{\textrm{P}}= \frac{1}{c^2}\dot{\mathcal{A}}_i^*\dot{\mathcal{A}}_i -\partial_i\mathcal{A}_j^*\partial_i\mathcal{A}_j+\partial_i\mathcal{A}_j^*\partial_j\mathcal{A}_i-\frac{\mu^2c^2}{\hbar^2}\mathcal{A}_i^*\mathcal{A}_i -\frac{1}{c}\left(\dot{\mathcal{A}}^*_i\partial_i\mathcal{A}_0 
+\partial_i\mathcal{A}_0^*\dot{\mathcal{A}}_i\right)+\partial_i\mathcal{A}_0^*\partial_i\mathcal{A}_0 +\frac{\mu^2c^2}{\hbar^2}\mathcal{A}_0^*\mathcal{A}_0. 
\end{equation}
This decomposition makes explicit the distinct roles of the temporal and spatial components. In particular, $\mathcal{A}_0$ appears without a time derivative and is therefore nondynamical. Instead, it is determined by a constraint involving the spatial components. Varying Eq.~\eqref{eq.Proca.expanded} with respect to $\mathcal{A}_0^*$ gives
\begin{equation}\label{eq.constraint}
-\frac{1}{c}\partial_i\dot{\mathcal{A}}_i+\Delta \mathcal{A}_0-\frac{\mu^2c^2}{\hbar^2}\mathcal{A}_0=0,
\end{equation}
which can be solved for $\mathcal{A}_0$ in terms of the spatial components $\mathcal{A}_i$. Substituting the solution of Eq.~(\ref{eq.constraint}) back into Eq.~(\ref{eq.Proca.expanded}) yields the reduced Lagrangian
\begin{equation}\label{eq.Proca.expanded2}
\mathcal{L}_{\textrm{P}}= \frac{1}{c^2}\dot{\mathcal{A}}_i^*\dot{\mathcal{A}}_i -\partial_i\mathcal{A}_j^*\partial_i\mathcal{A}_j+\partial_i\mathcal{A}_j^*\partial_j\mathcal{A}_i-\frac{\mu^2c^2}{\hbar^2}\mathcal{A}_i^*\mathcal{A}_i -\frac{1}{c^2}\partial_i\dot{\mathcal{A}}^*_i\left(\frac{\mu^2c^2}{\hbar^2}-\Delta\right)^{-1}\partial_j\dot{\mathcal{A}}_j.
\end{equation}
The resulting theory contains the three physical degrees of freedom of the massive vector field. No approximation has been made, and the reduced formulation is exactly equivalent to the original Proca theory.

We can now investigate whether these three components possess an internal $U(3)$ symmetry. More precisely, we consider transformations of the form
\begin{equation}
\mathcal{A}_i(t,\vec{x}) \mapsto \hat{U}_{ij}\mathcal{A}_j(t,\vec{x}),
\quad
\hat{U}\in U(3),
\end{equation}
where $\hat{U}_{ij}$ acts on the component index while leaving the spatial coordinates unchanged. Such a transformation would treat the three spatial components as an internal triplet, in analogy with a multiplet of three independent complex fields. The reduced Lagrangian in Eq.~(\ref{eq.Proca.expanded2}) does not make the answer immediately transparent. Some of its terms are not separately invariant under a general $U(3)$ transformation, but this alone does not establish that the complete action is not invariant.
The appropriate criterion is therefore to examine the complete quadratic operator governing the dynamics.

This can be done most conveniently in Fourier space. To proceed, we decompose the spatial components of the vector field as
\begin{equation}
\mathcal{A}_i(t,\vec{x})
=
\int\frac{d\omega\,d^3 k}{(2\pi)^{2}}
\,
a_i(\omega,\vec{k})
e^{-i(\omega t-\vec{k}\cdot\vec{x})}.
\end{equation}
Since the Proca theory is quadratic in the field, the action can then be written as
\begin{equation}
\mathcal{S}_{\rm P}
=
\int d\omega\, d^3k\,
a_i^*(\omega,\vec{k})
K_{ij}(\omega,\vec{k})
a_j(\omega,\vec{k}),
\end{equation}
where the momentum-space kernel is
\begin{equation}\label{eq.kernel}
K_{ij}(\omega,\vec{k})
=
\left(
\frac{\omega^2-\omega^2(k)}{c^2}
\right)\delta_{ij}
-
\left(
\frac{\omega^2-\omega^2(k)}
{\omega^2(k)}
\right)k_i k_j ,
\end{equation}
and the relativistic dispersion $\omega(k)$ relation is given by
\begin{equation}\label{eq.dispersion}
\omega(k)
=
\sqrt{\frac{\mu^2c^4}{\hbar^2}+c^2k^2} .
\end{equation}

The structure of the kernel makes the absence of an internal $U(3)$ symmetry manifest. Indeed, under a transformation
\begin{equation}
a_i(\omega,\vec{k})
\mapsto
\hat{U}_{ij}a_j(\omega,\vec{k}),
\quad
\hat{U}\in U(3),
\end{equation}
invariance of the action for arbitrary $\hat{U}_{ij}$ would require
\begin{equation}
\hat{U}^\dagger K(\omega,\vec{k})\hat{U}
=
K(\omega,\vec{k})
\end{equation}
for every $\omega$ and $\vec{k}$. The term proportional to $\delta_{ij}$ satisfies this condition, since $\delta_{ij}$ is invariant under $U(3)$. The second term, however, contains the matrix $k_i k_j$, 
which is not invariant under arbitrary internal rotations of the components.

The origin of this symmetry breaking is the coupling between the vector components and the spatial momentum. The index $i$ labels a spatial vector component rather than an independent internal degree of freedom, and the relativistic dynamics therefore distinguishes the component parallel to $\vec{k}$ from those orthogonal to it. An internal $U(3)$ transformation, by contrast, rotates the vector components while leaving the spatial momentum unchanged.\footnote{Notice that this does not contradict ordinary spatial rotational invariance. Under a spatial rotation, both the field components and the momentum vector transform, so the tensor $k_i k_j$ transforms covariantly. An internal $U(3)$ transformation is different: it acts only on the three field components while leaving the spatial coordinates, and hence $\vec{k}$, unchanged. It is precisely this distinction that allows the relativistic theory to be rotationally invariant while lacking the larger internal $U(3)$ symmetry.}  The $U(3)$ symmetry is consequently incompatible with the exact relativistic structure of the Proca theory.\footnote{It is also instructive to view the origin of the symmetry breaking in terms of the physical polarizations of the vector field. Introducing the transverse and longitudinal projectors,
\begin{equation}
P^{\rm T}_{ij}=
\delta_{ij}-\frac{k_i k_j}{k^2},
\qquad
P^{\rm L}_{ij}=
\frac{k_i k_j}{k^2},
\end{equation}
Eq.~(\ref{eq.kernel}) can be written as
\begin{equation}\label{eq.kernel.projectors}
K_{ij}
=
\frac{\omega^2-\omega^2(k)}{c^2}
P^{\rm T}_{ij}
+
\frac{\mu^2}{\hbar^2}
\frac{\omega^2-\omega^2(k)}{\omega^2(k)}
P^{\rm L}_{ij}.
\end{equation}
This decomposition makes explicit that the relativistic dynamics distinguishes the longitudinal polarization from the two transverse polarizations. The three physical polarizations therefore do not enter the quadratic action on an equal footing, providing a direct physical manifestation of the absence of an internal $U(3)$ symmetry.}

We can now use the structure of the kernel to determine how the internal symmetry changes in the nonrelativistic regime. This corresponds to the low-momentum limit
\begin{equation}
c^2k^2\ll \frac{\mu^2c^4}{\hbar^2},
\end{equation}
in which the relativistic dispersion relation~(\ref{eq.dispersion}) is dominated by the rest energy and takes the familiar nonrelativistic form
\begin{equation}
\omega(k)
=
\frac{\mu c^2}{\hbar}
+\omega_{\rm NR},
\qquad
\omega_{\rm NR}\simeq\frac{\hbar k^2}{2\mu}.
\end{equation}
After some trivial algebra, the momentum-space kernel~(\ref{eq.kernel}) can be written as
\begin{equation}\label{eq.kernel.NR.form}
K_{ij}(\omega,\vec{k})
=
\left(
\frac{\omega^2-\omega^2(k)}{c^2}
\right)
\left[
\delta_{ij}
-
\frac{c^2 k_i k_j}{\omega^2(k)}
\right].
\end{equation}
Note that the relative magnitude of the symmetry breaking term is controlled by
\begin{equation}
\frac{c^2k^2}{\omega^2(k)}
\sim
\frac{v^2}{c^2},
\end{equation}
so that, in the nonrelativistic regime,
\begin{equation}
K_{ij}(\omega,\vec{k})
=
\left(
\frac{\omega^2-\omega^2(k)}{c^2}
\right)
\left[
\delta_{ij}
+
\mathcal{O}\left(\frac{v^2}{c^2}\right)
\right].
\end{equation}
Thus, at leading nonrelativistic order, the three spatial components enter the quadratic operator on an equal footing, while the relativistic corrections distinguish the longitudinal polarization from the two transverse ones. The $U(3)$ symmetry is therefore an emergent symmetry of the leading-order nonrelativistic theory, broken by relativistic corrections at order $v^2/c^2$.

%%%%%%%%%%%%%%%%%%%%%%%%%%%%%%%%%%%%%%%%%%%%%%%%%%%% 
\section{Numerical evolution setup}\label{app.evol}
%%%%%%%%%%%%%%%%%%%%%%%%%%%%%%%%%%%%%%%%%%%%%%%%%%%%

In this appendix, we briefly summarize the numerical framework used to perform and analyze the time evolutions of the complex Einstein-Proca system presented in Sec.~\ref{sec:correspondece}.

\begin{figure}[t!]
\centering
\includegraphics[width=\textwidth]{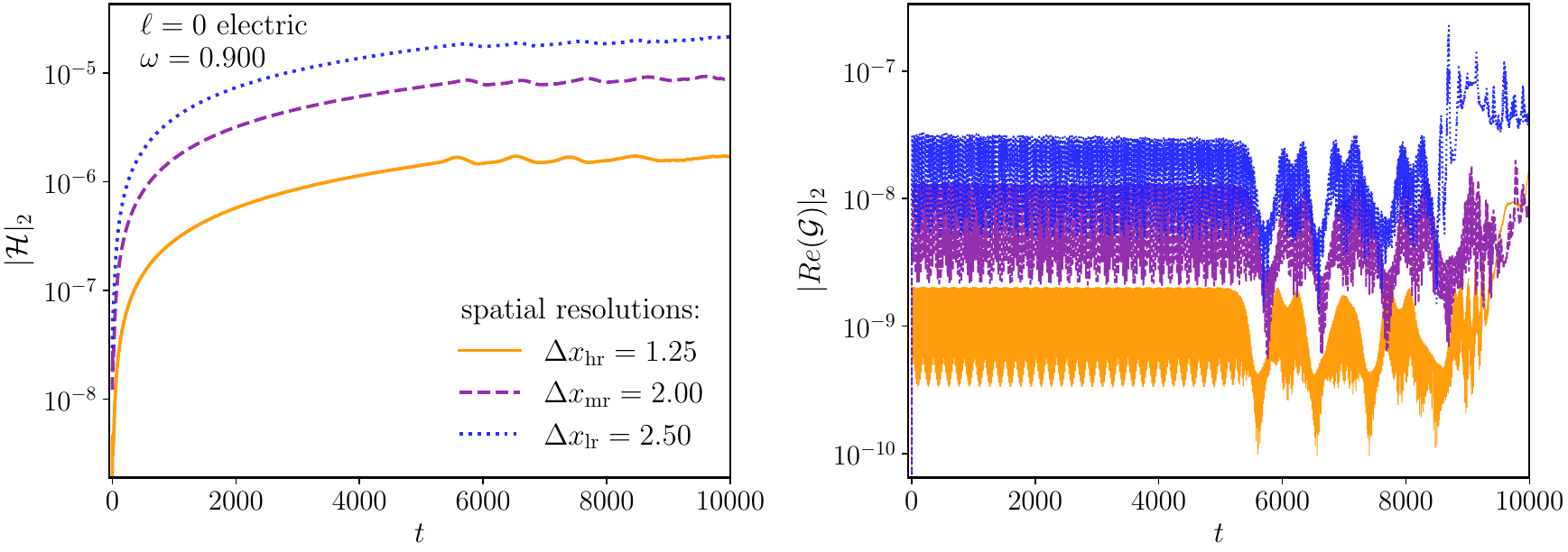}
\caption{\small{\textbf{Convergence analysis.} Global convergence test for the evolution of an $\ell=0$ electric Proca star with frequency $\omega=0.900$. {\it Left panel:} Time evolution of the $L_2$-norm of the Hamiltonian constraint for three different numerical resolutions. The relative scaling between the curves is consistent with an observed convergence order of approximately $3.6$. {\it Right panel:} Same as in left panel but for the $L_2$-norm of the Gauss constraint $\mathrm{Re}(\mathcal{G})$.}}
\label{fig:convergence}
\end{figure}

Following the standard 3+1 decomposition~\cite{Zilhao:2015tya,Sanchis-Gual:2017bhw}, the complex Proca field is split into components parallel and orthogonal to the future-directed timelike unit normal $n^\mu$ to the spatial hypersurfaces. To this end, one introduces the normal and spatial projections\footnote{Note that $\phi$ is not a spacetime component but only a name.}
\begin{equation}
\mathcal{X}_{\phi} := -n^{\nu}\mathcal{A}_{\nu},\quad 
\mathcal{X}_{\mu} := \gamma^{\nu}{}_{\mu}\mathcal{A}_{\nu},
\end{equation}
where $\gamma^{\mu}{}_{\nu}=\delta^{\mu}{}_{\nu}+n^{\mu}n_{\nu}$ is the spatial projection operator associated with the induced spatial metric $\gamma_{ij}$. The Proca field can then be reconstructed from its projections as
\begin{equation}
\mathcal{A}_{\mu} = \mathcal{X}_{\mu} + n_{\mu}\mathcal{X}_{\phi}.
\end{equation}
The spatial components of $\mathcal{X}_{\mu}$, given by $\mathcal{X}_{i} = \gamma^{\nu}{}_{i}\mathcal{A}_{\nu}$, together with $\mathcal{X}_{\phi}$, play the role of the vector and scalar potentials of the Proca field, respectively. The associated electric and magnetic fields are defined by
\begin{equation}
E_{i} : = \gamma^{\mu}{}_{i}\mathcal{F}_{\mu\nu}n^{\nu}, \qquad B_{i} : = \gamma^{\mu}{}_{i}\,{}^{\star}\mathcal{F}_{\mu\nu}n^{\nu} = \epsilon_{ijk} D^{j}\mathcal{X}^{k},
\end{equation}
and both fields are purely spatial, satisfying $E_{\mu}n^{\mu}=B_{\mu}n^{\mu}=0$. Here $D_i$ denotes the covariant derivative compatible with the spatial metric $\gamma_{ij}$, $\epsilon_{ijk}$ is the spatial Levi-Civita tensor, and $^{\star}\mathcal{F}_{\mu\nu}:=\frac{1}{2}\epsilon_{\mu\nu\alpha\beta}\mathcal{F}^{\alpha\beta}$ is the Hodge dual of the Proca field strength tensor.

The evolution system is supplemented by the Hamiltonian, momentum, and Gauss constraints,
\begin{subequations}
\begin{eqnarray}
\mathcal{H} &=& R - K_{ij}K^{ij} + K^{2} - 2\left(E^{i}E_{i}+B^{i}B_{i} +\mu^{2}\left(\mathcal{X}_{\phi}^{2} +\mathcal{X}^{i}\mathcal{X}_{i}\right)\right)=0, \\
\mathcal{M}_{i} &=& D^{j}K_{ij}-D_{i}K -2\left(\epsilon_{ijk}E^{j}B^{k} +\mu^{2}\mathcal{X}_{\phi}\mathcal{X}_{i}\right)=0, \\
\mathcal{G} &=& D_{i}E^{i}+\mu^{2}\mathcal{X}_{\phi}=0.
\end{eqnarray}
\end{subequations}
The simulations were performed with the \texttt{Einstein Toolkit}~\cite{EinsteinToolkit:2024_11, loffler2012einstein}, based on the \texttt{Cactus} infrastructure~\cite{Goodale:2002a} with mesh refinement. The spacetime evolution is carried out with the \texttt{McLachlan} code~\cite{brown2009turduckening, reisswig2011gravitational}, which implements the Baumgarte-Shapiro-Shibata-Nakamura formulation, while the Proca sector follows the framework developed in~\cite{Zilhao:2015tya, witek_2023_7791842, Sanchis-Gual:2018oui, Sanchis-Gual:2022mkk, Herdeiro:2023wqf, Lazarte:2025wlw}. Time integration is performed using the method of lines with a fourth-order Runge-Kutta scheme.

For the evolutions reported in Sec.~\ref{sec:correspondece}, we use three nested refinement levels without imposing symmetries. Configurations with $\omega \le 0.970$ are evolved using box sizes $(99,49.5,24.75)$, and corresponding grid spacings $(1.10,0.55,0.275)$. For configurations with $\omega \ge 0.980$, we instead employ box sizes $(110,55,27.5)$, with grid spacings $(0.88,0.44,0.22)$. In both cases, the first set specifies the size of each refinement level, while the second gives the corresponding spatial resolution. 
Successive refinement levels increase the spatial resolution by a factor of two toward the central region. All quantities are expressed in terms of the relativistic code variables introduced in Eq.~(\ref{eq.dimensionless.relativistic}).

We have performed a convergence study, reported in Fig.~\ref{fig:convergence}, using a slightly different refinement setup with nested box sizes $(110,55,27.5)$, and grid spacings refined by a factor of two between successive refinement levels. Three different overall resolutions were considered, corresponding to coarsest-level grid spacings $\Delta x_{\rm hr}=1.25$, $\Delta x_{\rm mr}=2.00$, and  $\Delta x_{\rm lr}=2.50$. The results exhibit a convergence order consistent with the fourth-order accuracy of the evolution scheme, supporting the interpretation that the instability observed in the $\ell=0$ electric Proca stars is physical rather than a numerical artifact.

\begin{figure}[t!]
\centering
\includegraphics[width=1.0\textwidth]{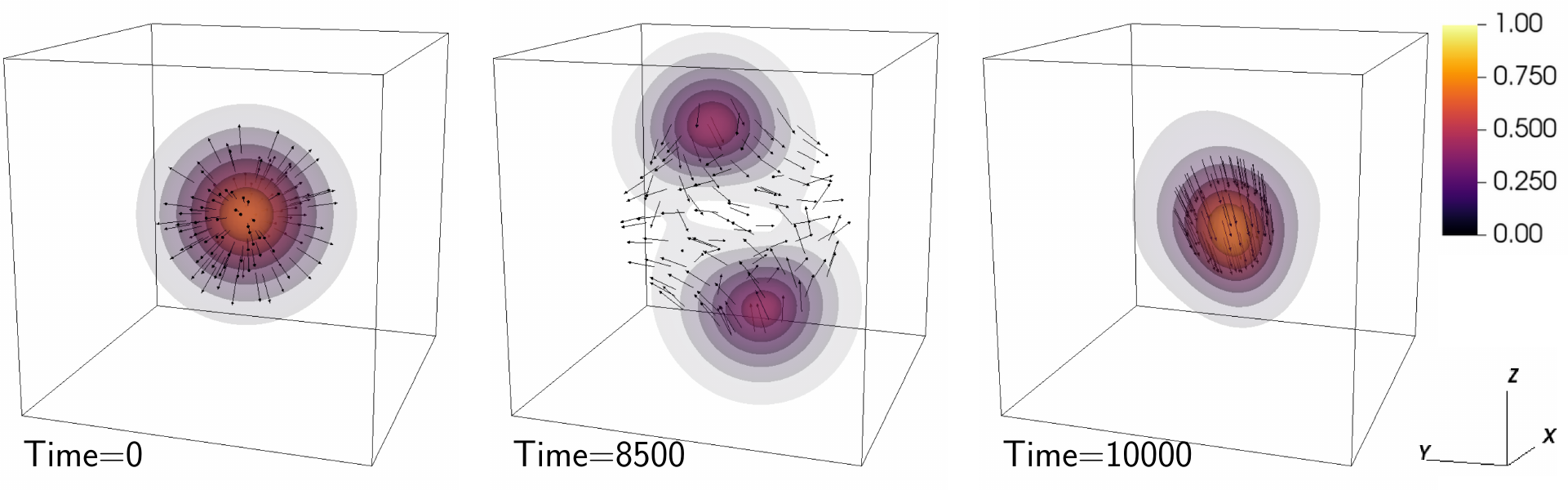}
\caption{\small{\textbf{Dynamical migration from the $\ell=0$ to the $\ell=1$ branch.}} Snapshots of the time evolution of a relativistic, nodeless $\ell=0$ electric Proca star with $\omega=0.900$. Arrows show the real part of the spatial Proca field, $\mathrm{Re}(\mathcal{X}_i)$, while the color shading indicates the energy density normalized to its maximum value during the evolution. The polarization vector provides a direct visualization of the transition from the initial radially polarized configuration toward a state with quasi-linear polarization.}
\label{fig:3d_evovolution}
\end{figure}

The evolutions show that the unstable configurations dynamically migrate toward an $\ell=1$-like state. This process takes place in two distinct stages, which can be clearly identified in the right panel of Fig.~\ref{fig:stability} for the configuration with $\omega=0.950$. During the first stage, beginning at $t\sim 4150$, the spherical symmetry of the system is broken due to the nonlinear effects of a growing unstable mode, and the initially shell-like configuration evolves into a prolate dipolar structure characterized by two symmetric lobes. At this stage, the  $\mathbb{Z}_2$ reflection symmetry along the symmetry axis is still preserved. In the second stage, which develops on a longer timescale and is associated with nonlinear dynamics along the $z$-direction, occurring approximately at $t\sim 11200$, this  $\mathbb{Z}_2$ symmetry is broken, and one of the lobes is progressively dispersed while the other becomes dominant, signaling the decay toward a single-lobed prolate configuration.

This behavior is further illustrated in Fig.~\ref{fig:3d_evovolution}, where for the configuration with $\omega=0.900$ the vector potential evolves from a radial polarization pattern to a quasi-linearly polarized configuration. The combined morphological and polarization evolution provides additional evidence that the end state solution corresponds to a relativistic $\ell=1$ electric Proca star.

\begin{figure}[t!]
\centering
\includegraphics[width=\textwidth]{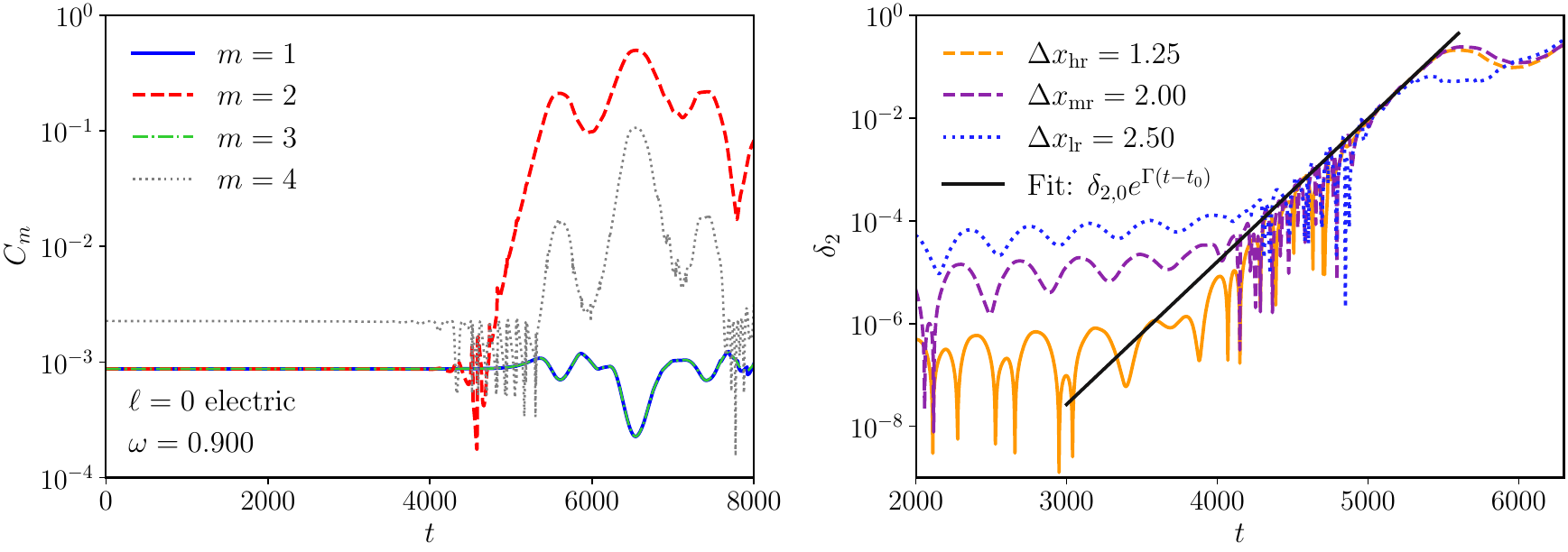}
\caption{\small{\textbf{Analysis of the unstable mode.} Evolution of the azimuthal Fourier modes of the ADM energy density for the $\ell=0$ Proca star with $\omega=0.900$, shown on a logarithmic scale. {\it Left panel:} The $m=2$ mode is the dominant unstable mode. {\it Right panel:} Exponential growth of the $m=2$ mode and corresponding fit, yielding essentially the same growth rate at the three resolutions. For visual clarity, the low- and medium-resolution data have been shifted in time relative to the high-resolution run so that the curves overlap during the exponential-growth regime.}}
\label{fig:unstable_mode}
\end{figure}

The unstable mode responsible for the initial stage of this migration can be identified through the azimuthal Fourier decomposition of the ADM energy density $\rho_{\rm ADM}$ (see~\cite{Lazarte:2025wlw} for details), defined as
\begin{eqnarray}
    C_{m}(t) = \int \text{d}^3x\sqrt{\gamma}\,\rho_{\rm ADM}(t,\vec{x})\,e^{im\varphi},
\end{eqnarray}
where $\gamma$ is the determinant of the spatial metric and $\varphi$ is the azimuthal angle in a cylindrical coordinate system centered on the star's center of mass. For the dynamical migration of the configuration with $\omega=0.900$, we computed the modes $m = 1, 2, 3, 4$, and found that the $m=2$ mode is dominant and grows exponentially. All four modes are shown in the left panel of Fig.~\ref{fig:unstable_mode} for the highest-resolution run used in the convergence study described above. To isolate the exponential growth of $C_2$ from the background, we subtract its initial value, defining
\begin{equation}
    \delta_2 := |C_2(t) - C_2(t=0)|.
\end{equation}
We fit the exponential-growth regime of $\delta_2$ to $\delta_2(t)=\delta_{2,0}\,e^{\Gamma(t - t_0)}$, obtaining a growth rate $\Gamma\approx0.0064$. This quantity is an intrinsic property of the unstable configuration and allows us to define the lifetime between two specified amplitudes as
\begin{equation}
t_{\rm life}=\frac{1}{\Gamma}\ln\left(\frac{\delta_{2,\rm fin}}{\delta_{2,\rm ini}}\right),
\end{equation}
in addition to the lifetime defined from the macroscopic evolution of Fig.~\ref{fig:stability}. The evolution of $\delta_2$ for the three resolutions used in the convergence study, together with the same exponential fit, is shown on a logarithmic scale in the right panel of Fig.~\ref{fig:unstable_mode}. The nearly identical slopes in the exponential-growth regime yield essentially the same value of $\Gamma$ at all three resolutions, supporting the interpretation that the observed growth is governed by the same physical unstable mode rather than by numerical artifacts. As a consistency check between the two lifetime estimates, we use the numerical floor of the high-resolution simulation, $\delta_{2,\rm ini}\sim10^{-7}$, and the characteristic amplitude at the onset of the nonlinear regime, $\delta_{2,\rm fin}\sim10^{-1}$, to obtain $t_{\rm life}\sim 2158$.
This is in excellent agreement with the value $2047$ reported in Fig.~\ref{fig:stability} for $\omega=0.900$.

%%%%%%%%%%%%%%%%%%%%%%%%%%%%%%%%%%%%
\bibliographystyle{jhep}  
\bibliography{references}

\end{document}